\documentclass[%
a4paper,
aps,
prb,
onecolumn,
nofootinbib,
amsmath,amssymb,
superscriptaddress,
10pt,
]{revtex4-2}

\usepackage[a4paper,centering,hmargin=2.25cm,vmargin=1.75cm]{geometry}

\usepackage{graphicx}
\usepackage{amsfonts,amsmath}
\usepackage{mathtools}
\usepackage{xcolor}
\usepackage[colorlinks=true,citecolor=blue,linkcolor=blue,urlcolor=blue]{hyperref}
\usepackage{physics}
\usepackage{braket}
\usepackage{ulem}
\usepackage[capitalize]{cleveref}

\usepackage{microtype}

\let\emph\relax
\DeclareTextFontCommand{\emph}{\itshape}

\def\rank{\mathcal{R}}
\newcommand{\Fig}[1]{{Fig.~{\ref{#1}}}}

\newcommand{\beq}{\begin{equation}}
\newcommand{\eeq}{\end{equation}}

\begin{document}

\title{Spatio-temporal tensor-network approaches to out-of-equilibrium dynamics bridging  open and closed systems}% Force line breaks with \\

\author{Sergio Cerezo-Roquebrún}
\affiliation{Instituto de Física Teórica UAM/CSIC, C/ Nicolás Cabrera 13--15, Cantoblanco, 28049 Madrid, Spain}

\author{Aleix Bou-Comas}
\affiliation{Institute of Fundamental Physics IFF-CSIC, C/ Serrano 113b, Madrid 28006, Spain}

\author{Jan T.\ Schneider}
\affiliation{Institute of Fundamental Physics IFF-CSIC, C/ Serrano 113b, Madrid 28006, Spain}

\author{Esperanza López}
\affiliation{Instituto de Física Teórica UAM/CSIC, C/ Nicolás Cabrera 13--15, Cantoblanco, 28049 Madrid, Spain}

\author{Luca Tagliacozzo}
\affiliation{Institute of Fundamental Physics IFF-CSIC, C/ Serrano 113b, Madrid 28006, Spain}
% \email{luca.tagliacozzo@iff.csic.es}

\author{Stefano Carignano}
\email{stefano.carignano@bsc.es}
\affiliation{Barcelona Supercomputing Center, 08034 Barcelona, Spain}

\date{\today}

\begin{abstract}
% abstract text!
The study of many-body quantum systems out of equilibrium remains a significant challenge with complexity barriers arising in both state and operator-based representations.
In this work, we review recent approaches based on finding better contraction strategies for the full spatio-temporal tensor networks that encode the path integral of the dynamics, as well as the conceptual integration of influence functionals, process tensors, and transfer matrices within the tensor network formalism.
We discuss recent algorithmic developments, highlight the complexity of influence functionals in various dynamical regimes and present consistent results of different communities, showing how ergodic dynamics render these functionals exponentially difficult to compress.
Finally, we provide an outlook on strategies to encode complementary influence functional overlaps, paving the way for accurate descriptions of open and closed quantum systems with tensor networks.

\end{abstract}

\maketitle

% \tableofcontents

\section{Introduction}\label{sec:intro}
Many-body systems out-of-equilibrium still defy our understanding, and 
our intuition of the origin of their complexity is still evolving.
In the case of closed quantum systems, where the  many-body system is ideally isolated from the rest of the world constituting its environment, the dynamics is governed by the unitary evolution dictated by the Schrödinger equation.
Such a unitary evolution can be applied to the state of the system or to its operators.
This simple fact leads to different pictures about the difficulty of solving the dynamics.
If the evolution is applied on states, in the Schrödinger picture, they become increasingly complex, and simple tensor networks ansätze struggle to describe them with polynomial resources, something that is known as the entanglement barrier~\cite{calabrese_2005,lauchli_spreading_2008,dubail2017}.

Contrarily, if the evolution is applied to operators, in the Heisenberg picture, there are specific forms of evolution that can be solved.
For example, for spin--1/2 systems, Clifford circuits map Pauli strings into themselves, and thus the dynamics can be efficiently described~\cite{PhysRevA.57.127}.
Additionally, the dynamics governed by integrable Hamiltonians are conjectured to generate only a small amount of local operator entanglement and thus describing the evolution of local operators in such systems has the same complexity of describing a local quench on states, allowing for an efficient description~\cite{prosenEfficiencyClassicalSimulations2007,prosen2007operator,bertiniOperatorEntanglementLocal2020,bertiniOperatorEntanglementLocal2020a,Giudice_PhysRevLett.128.220401, Thoenniss_PhysRevB.107.195101}.
However, generic interacting systems generate again a barrier of operator entanglement and thus suffer from the same shortcomings as the simulation of the evolution of the states.

Given the complexity barriers in both state and operator representations, a growing trend is to adopt an open-system perspective. Instead of attempting to describe the full system dynamics, this approach focuses on the evolution of few-body correlation functions\cite{banuls2009,muller-hermes2012,surace2019a,white2018,frias2022,paeckel2019time}. This perspective naturally connects the dynamics of closed quantum systems to those of open systems. 
From the open-system viewpoint, the few bodies involved in the correlation functions define the \emph{system}, while the remainder of the many-body system acts as the \emph{environment}.

%The traditional distinction between open-system dynamics and closed-system dynamics, is thus fading away.
Traditionally, significant progress in understanding open-system dynamics has been made by studying a small subsystem, such as an atom in a cavity or an impurity in a metal, and describing it with master equations. While the full system and environment evolve unitarily, the subsystem undergoes dissipative evolution driven by a trace-preserving quantum channel. This framework provides a significant simplification compared to the full system's description, as the subsystem often has a finite-dimensional Hilbert space, allowing the quantum channel to map between finite-dimensional density matrices. However, obtaining such master equations has historically relied on analytical approximations, such as weak coupling or memory-less environments, defining Markovian systems where the current state suffices to predict future states. Even in this context, the precise definition of Markovianity remains a subject of debate as seen by comparing for example the definition in~\cite{rivas2010} with the one in~\cite{dowling2024}.

This work reviews recent advances in designing tensor network algorithms to study out-of-equilibrium dynamics. These algorithms unify the descriptions of closed and open dynamics with minimal approximations, particularly for one-dimensional many-body systems. The starting point of these approaches are the spatio-temporal tensor networks that encode the path integral of the dynamics. Following the original Feynman--Vernon idea~\cite{feynman1963}, we define influence functionals  once we identify a region as the system while the rest constitutes the environment. Partial integration of the path integral over the spatio-temporal degrees of freedom of the environment gives rise to the influence functionals.
Furthermore, in this framework it is easy to show how these influence functionals naturally emerge as partial traces of so-called process tensors,
%, partial contractions of spatio-temporal patches of the path integral.
%Process tensors,
which were initially introduced in the quantum information community \cite{chiribella2008} to generalize quantum channels and study the role of specific gates in quantum circuits.
 % here play thus an important role.
%

We also show how these distinct concepts naturally integrate within the spatio-temporal tensor network framework. Additionally, finding the correct influence functional corresponds to solving an open-system dynamics problem, where spatial transfer matrices act as quantum channels driving the dissipative evolution of influence functionals. Specifically, the  transfer matrices evolve the influence functionals in space, that is, they generate the influence functional of a smaller system (larger environment) from that of a larger region, or that of a smaller environment by incorporating new sites into the environment. This paper briefly outlines the algorithms developed to achieve this task.
After introducing all these objects and the algorithms that we practically use to compress them in simple tensor networks,
we  also review the known results about the complexity of the influence functionals for different classes of dynamics. We will thus unveil how for generic ergodic dynamics, influence functionals are exponentially hard to compress.
We conclude with an outlook on the potential to accurately describe the dynamics of open and closed quantum systems using tensor networks. This involves attempting to directly encode the overlaps of complementary influence functionals (e.g., left and right) rather than the individual influence functionals separately.

\section{Tensor network approach to the influence functional}\label{sec:2}

In this section, we will see how the concept of influence functionals (IFs), introduced to describe how the environment affects a system within the path integral formulation of quantum mechanics~\cite{feynman1963}, can be represented using tensor networks (TNs).
The connection is made possible by the ideas of \textit{transverse contraction} of the TN associated with the time evolution of a quantum system \cite{banuls2009, hastings2015, frias2022, Carignano_PhysRevResearch.6.033021} and of temporal matrix product states (tMPS), which can be used to describe these functionals.

To see this in more detail,
let us take as starting point the typical scenario of a quantum quench where one begins with a given initial state, usually a product state or the ground state of some Hamiltonian $H_0$, and evolves it under a Hamiltonian $H\neq H_0$. For the sake of simplicity, we will restrict ourselves to one-dimensional chains so that states can be typically expressed through a matrix product state (MPS) ansatz, although our results can be generalized to any number of spatial dimensions $D$.
Furthermore, we will assume systems whose constituents are described by Hilbert spaces of dimension $d$, and that interact only with nearest neighbors, namely $H=\sum_{i} h_{i,i+1}$, where the operators $h_{i,i+1}$ act only on the constituents $i$ and $i+1$.

When describing the dynamics of a one-dimensional closed system through tensor networks, one can 
integrate the Schrödinger equation by  constructing the time-evolution operator $U(T)=\exp(-iHT)$ and apply it on states or operators (see, e.g.,~\cite{paeckel2019time} for a review on time evolution methods using MPS). 

Since $U(T)$ is a highly non-local operator, some previous steps are needed to express it as a tensor network. 
One of the standard approaches \cite{vidal2004efficient} is to first split the evolution into small time steps of size $\delta t=T/n_T$, so that $U(T) = [U(\delta t)]^{n_T}$. 
Next, by using the so-called Suzuki-Trotter decomposition, the operator $U(\delta t)$ is approximated as a product of the 2-body gates $U_{i,i+1}\equiv\,\exp(-i h_{i,i+1}\delta t)$, at the cost of committing some error that depends on $\delta t$ as well as the non-commutativity of the $\{h_{i,i+1}\}$. 
For a system with nearest-neighbor interactions, this decomposition is typically done by dividing the Hamiltonian into operators acting on odd and even bonds, i.e., $H=H_{\mathrm{odd}}+H_{\mathrm{even}}$ where $H_{\mathrm{odd} (\mathrm{even})}=\sum_{i\in \mathrm{odd (even)}} h_{i,i+1}$, so that each term of $H_{\mathrm{even}}$ (or $H_{\mathrm{odd}}$) commute with each other as they act on different constituents. 

We can then build the first-order Suzuki-Trotter decomposition,
\beq
    e^{-iH\delta t} = e^{-iH_{\mathrm{odd}}\delta t} e^{-iH_{\mathrm{even}}\delta t} + \mathcal{O}(\delta t^2)\,,
\eeq
or higher-order approximations, with a Trotter error of order $\mathcal{O}(\delta t^{n+1})$ for an $n$-th order decomposition \cite{paeckel2019time}.

\begin{figure}[t]
    \centering
    \includegraphics[width=0.9\linewidth]{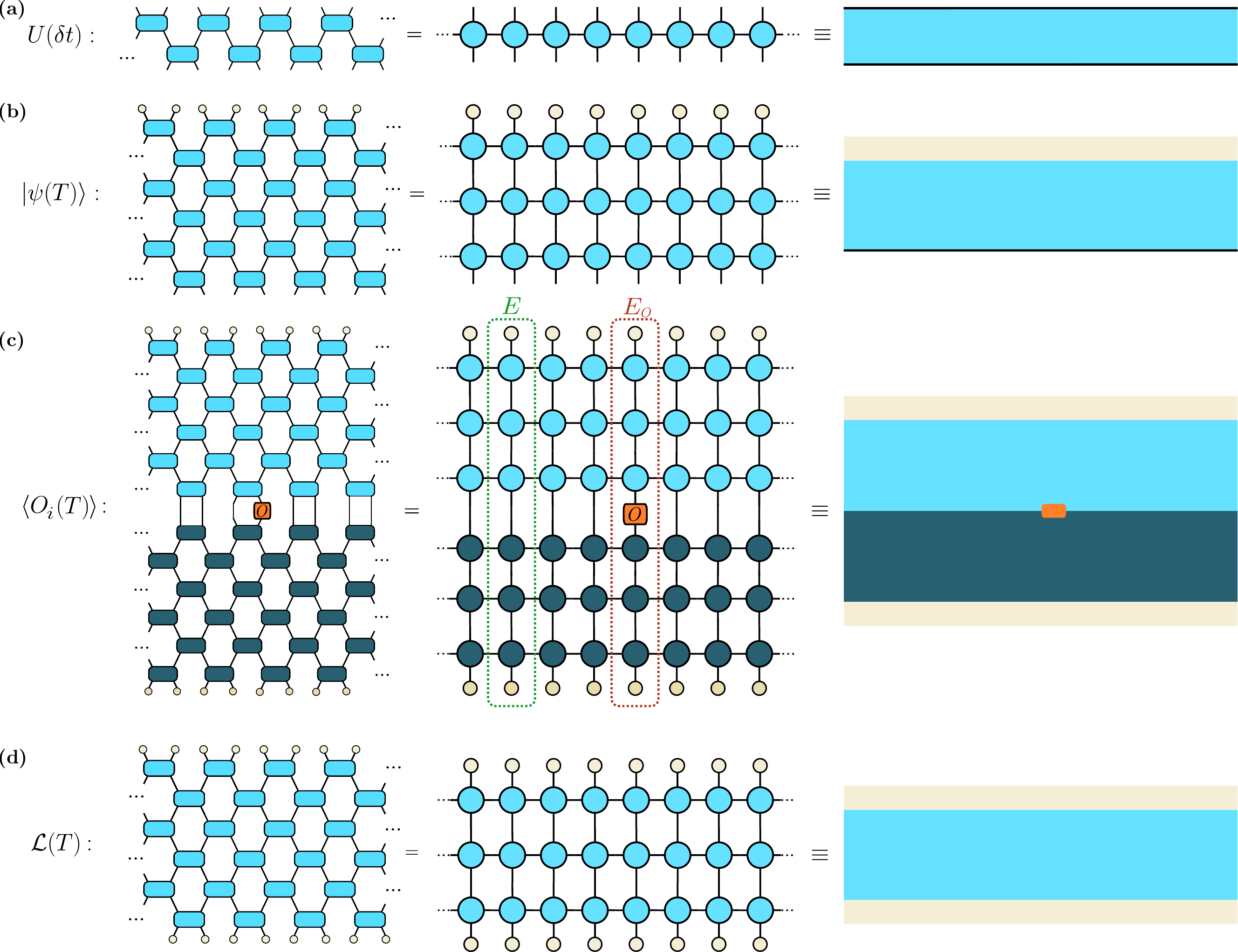}
    \caption{\label{fig:brickwall-tn-sheet}
    Time runs from top to bottom. All quantities are represented as (left column) a brick wall circuit, (center column) with a matrix product operators, and (right column) in an abstract representation, where the black boundaries denote open indices. 
    (a) A graphical representation of the first-order Suzuki--Trotter decomposition of the time evolution operator \(U(\delta t)\) for a small time step \(\delta t\) and for nearest-neighbor interactions.
    (b) \(\ket{\psi(T)} = \prod_{i=1}^{n_T} U(\delta t) \ket{\psi_0}\), the Trotterized time evolution of an initial product state \(\ket{\psi_0}\) (depicted in gray) up to time \(T = n_T \cdot \delta t\).
    (c) \(\braket{O_{i} (T)}\), the time-dependent expectation value of a single body observable \(\braket{O_{i} (T)} = \expval{{[U^\dagger(\delta t)]}^{n_T} \, O_{i} \,{[U(\delta t)]}^{n_T}}{\psi_0}\) (Darker shaded tensors refer to $U^\dagger$). We have highlighted the transfer matrix \(E(T)\) in green, and the operator transfer matrix \(E_O(T)\) in orange.
    (d) The Loschmidt echo \(\mathcal{L}(T) = \expval{{[U(\delta t)]}^{n_T}}{\psi_0}\).
    }
\end{figure}

\begin{figure}[t]
    \centering
    \includegraphics[width=0.95\linewidth]{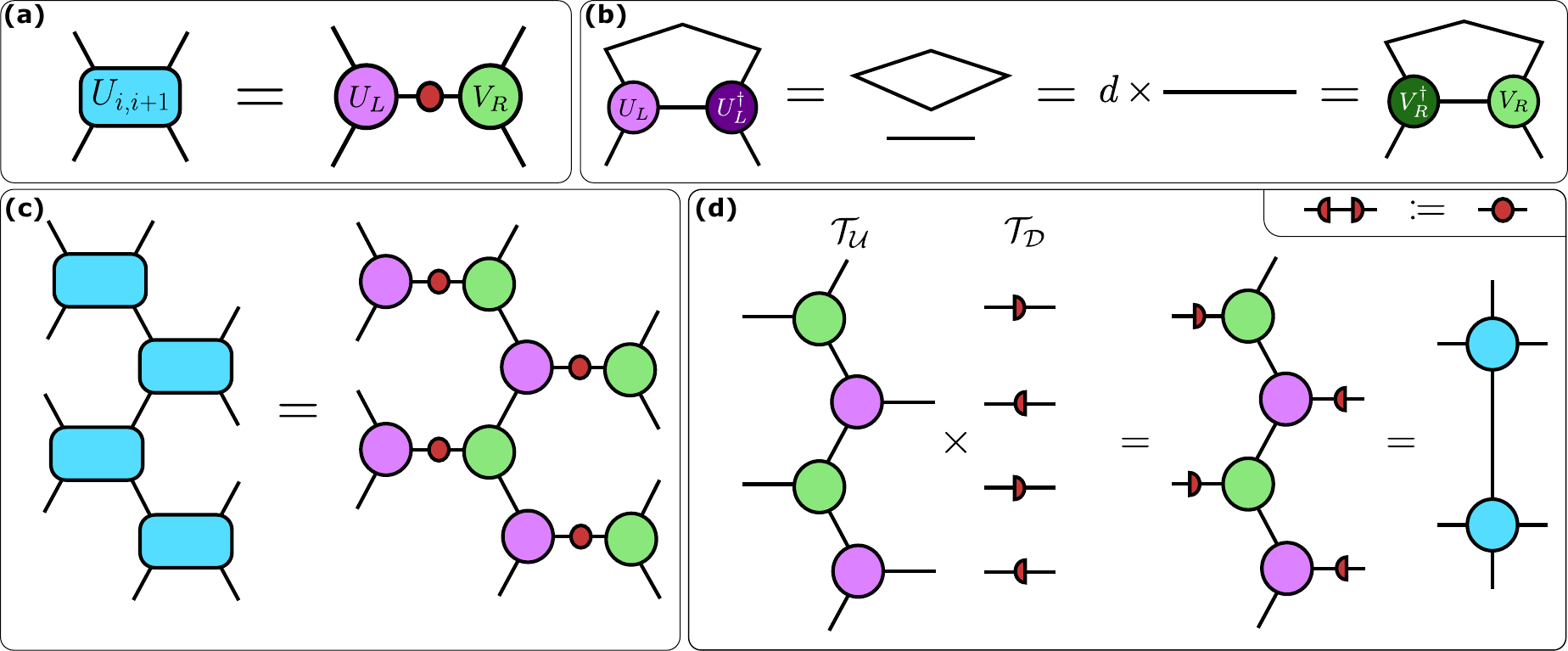}
    \caption{\label{fig:unitarity}%
        \textbf{(a)} By performing a singular value decomposition of elementary unitary gates derived from the Trotter expansion of the real time dynamics induced by the two-qudit time evolution operator \(U_{i,i+1} = U_L \cdot \Sigma \cdot V_R\). Panel \textbf{(b)} shows the normalization conditions of the relevant tensors. We can define the two-dimensional tensor network \textbf{(c)} and \textbf{(d)} describing the dynamics and interpret the evolution in space as a sequence of unitary evolutions driven by $\mathcal{T}_U$ plus weak measurements driven by $\mathcal{T}_D$.
    }
\end{figure}

By decomposing each $U(\delta t)$, we can express the entire evolution as a brick-wall circuit, as shown in the first column of \cref{fig:brickwall-tn-sheet}. 
Alternatively, one can go one step further by factorizing the two-body gates and suitably grouping the resulting tensors, building the matrix product operator (MPO) associated with $U(\delta t)$ (see \cref{fig:brickwall-tn-sheet}, center column). 
The decompositions required to build the MPO can be performed using, e.g., a singular value decomposition\footnote{Alternatively, one can decompose the $U_{i,i+1}$ in a basis of one-site operators, leading to a similar structure.}  (SVD), $U_{i,i+1}=U_L\,\Sigma\, V_R$, as depicted in \cref{fig:unitarity} (For later discussion, we will assume that $U_L$ and $V_R$ are square matrices, ensuring that they are proper unitaries).

As a result of the discretization of the time evolution, any dynamical quantity we are interested in will be expressed in terms of numerous tensors, giving rise to a two-dimensional network that has to be contracted.
To give an example, the TN representing the time-dependent expectation value of a local operator acting on a few sites of the system, $\langle O(T) \rangle$, is depicted in \Fig{fig:brickwall-tn-sheet}(c) and written from top to bottom as follows: starting from an initial state $\ket{\psi_0}$, we apply $U(T)$, then the operator $O$, followed by the Hermitian conjugate of the time-evolved state, $\bra{\psi_0} U^\dagger(T)$.
The latter can be seen as a part of ``backward'' time evolution, so that in the Keldysh formalism $\ket{\psi(T)} = U(T)\ket{\psi_0}$ represents the ``forward'' contour, while $\bra{\psi(T)}$ the ``return'' contour \cite{hastings2015, tirrito2018characterizing}, with the operator $O$ at the middle.
Another quantity of interest is the return amplitude of a time-evolved state to its initial configuration, which can be referred to as a Loschmidt echo (\cref{fig:brickwall-tn-sheet}(d)): 
\begin{equation}
\label{eq:loschmidt_echo}
{\mathcal L}(T) = \braket{\psi_0|\psi(T)} = \braket{\psi_0|U(T)|\psi_0} \,.
\end{equation}

The task of contracting two-dimensional TNs is, in general, exponentially hard \cite{schuch2007computational}.
The traditional prescriptions of quantum mechanics typically involve a contraction row by row along the temporal axis. In this case, the complexity of the contraction is dictated by the entanglement entropy of the time-evolved $\ket{\psi(T)}$ if the contraction starts from the initial state \cite{verstraete2006}, or by the operator entanglement if the contraction involves the Heisenberg evolution of an operator \cite{prosen2007operator, prosenEfficiencyClassicalSimulations2007, pizornOperatorSpaceEntanglement2009}.
Due to the entanglement barrier, these entropies typically increase linearly with time, corresponding to an exponential growth of the bond dimensions involved, so generally, these contractions cannot be carried out efficiently.
Yet, it is clear that another contraction direction is possible: one can namely start from the left and right edges of the system and perform contractions along the space direction~\cite{banuls2009, hastings2015, Carignano_PhysRevResearch.6.033021}.
The basic building blocks involved in this ``transverse'' contraction are now the columns of the 2D network, which can be seen as states and operators defined at different moments of time for a fixed spatial site. We will refer to them as ``temporal'' MPS and MPO (tMPS and tMPO), respectively.
The contraction procedure then goes as follows: starting from the edges of a system of $N$ constituents we identify a ``left'' tMPS $\bra{L^{[1]}}$ and a ``right'' one $\ket{R^{[N]}}$, as well as the $j$-th column tMPO $E^{[j]}(T)$, cf.\ \cref{fig:brickwall-tn-sheet}(c). 
At each step, we build the tMPS,
\begin{align}
    \bra{L^{[k]}}&\rightarrow \bra{L^{[k+1]}}=\bra{L^{[k]}}E^{[k+1]}(T)\,, \\
    \ket{R^{[l]}}&\rightarrow \ket{R^{[l-1]}}=E^{[l-1]}(T)\ket{R^{[l]}} \,,
\end{align}
by applying to them the columns associated with the neighboring site. In this way, the 2D TN equals the overlap between the two tMPS representing the transverse contraction until adjacent columns. Thus, quantities like the Loschmidt echo are given by $\mathcal{L}(T)=\braket{L^{[i]}_\mathcal{L}|R^{[i+1]}_\mathcal{L}}$, while for instance the expectation value of a local observable $O$ acting on the $i$-th site reads $\braket{O_i (T)}=\bra{L^{[i-1]}}E^{[i]}_O\ket{R^{[i+1]}}$, being $E^{[i]}_O$ the tMPO column with the insertion of the operator (see \cref{fig:brickwall-tn-sheet}(c)).
Analogously to what happens for MPS defined in space, we are interested in finding the relevant quantity dictating whether an efficient representation of these transverse states is possible, e.g., a ``temporal'' entanglement \cite{banuls2009, hastings2015, Carignano_PhysRevResearch.6.033021, lerose2023overcoming}, which can be seen as a measure of correlations of a subsystem with itself at different times. 
It is noteworthy that, while the transfer matrix evolving in time $U(\delta t)$ corresponds to a unitary operator, for the transverse contraction the spatial evolution is governed by the non-unitary $\{E^{[k]}\}$, as can be seen by the decomposition in \Fig{fig:unitarity}(c)--(d).
In this sense, the contrasting nature of the transverse evolution with respect to the usual one can give rise to a different behavior of the complexity of contracting the whole 2D TN with time, as reviewed in Sec.~\ref{sec:5}.
More specifically, from \cref{fig:unitarity}, we can see one step of space translation as a unitary evolution \(\mathcal{T}_U\) plus an additional insertion of real diagonal matrices \(\mathcal{T}_D\) containing singular values. After proper normalization, these operations can be related to weak measurements \cite{ippoliti2021}.

\begin{figure}[t]
    \centering
    \includegraphics[width=0.6\linewidth]{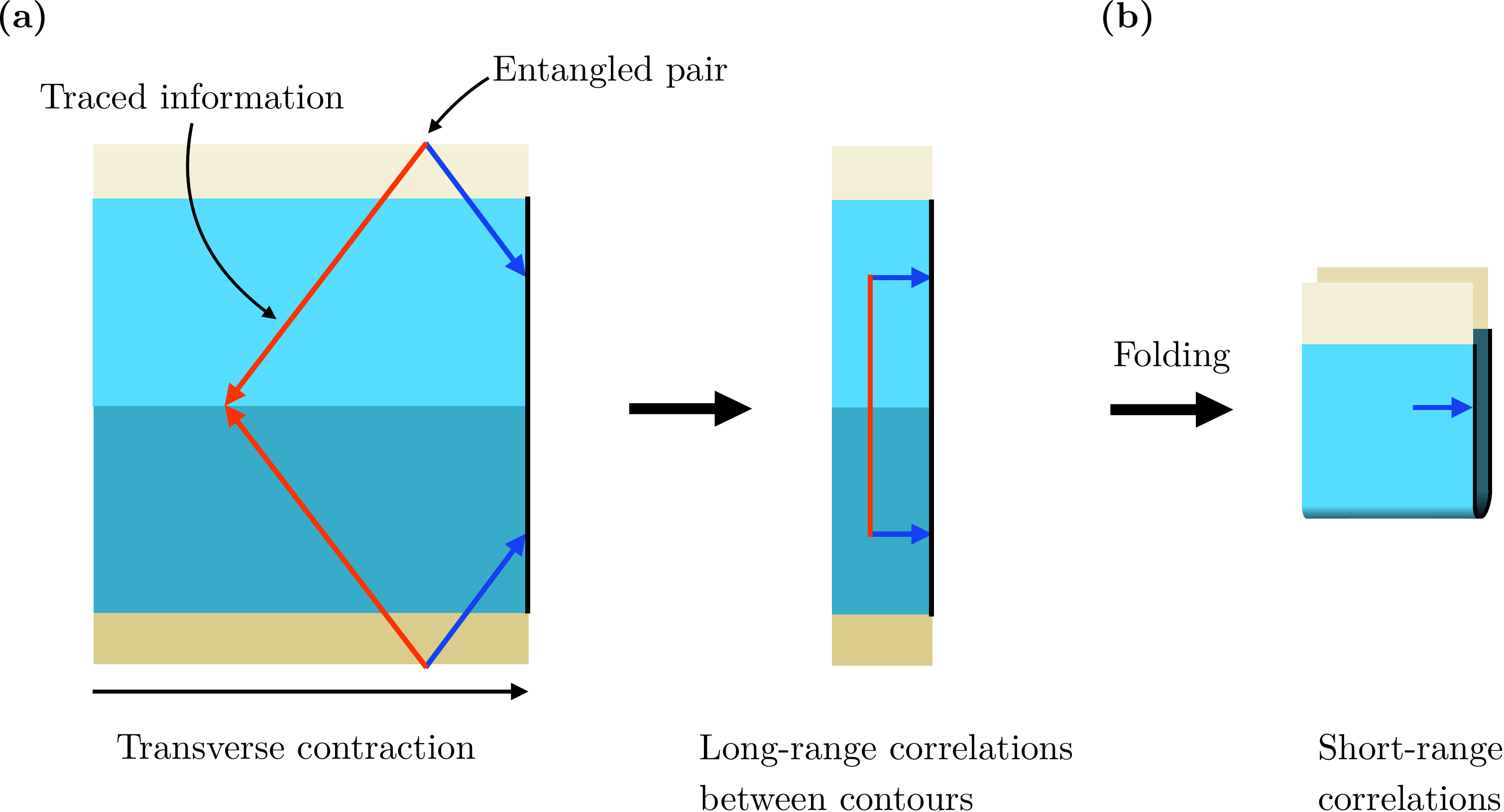}
    \caption{\label{fig:corr_contours}%
    For a setup including both forwards and backwards time evolution (cfr. \cref{fig:brickwall-tn-sheet}(c)), we show in 
    (a) a graphical representation of how the correlations between contours are formed for a local entangled pair that propagates freely in opposite directions.
    The excitation marked in red remains in the contracted subsystem throughout the entire evolution, so that it is traced in the transverse contraction. 
    On the other hand, the excitation marked in blue leaves the subsystem and enters into the complement. Due to the spatial entanglement between the red and blue excitations, the transverse contraction leads to correlations between contours. 
    (b) The associated long-range correlations are converted into short-range correlations through the folding operation.}
\end{figure}

%Concerning the efficiency of the transverse contraction,
Depending on the network structure associated with the desired dynamical quantity, further operations can be performed to carry out the transverse contraction in a more efficient way. In particular, if we are interested in computing the time evolution of the expectation values of a local operator or few-body correlators, the transverse contraction may induce correlations between the forward and return contours if there is information partially traced out (see \cref{fig:corr_contours}(a) for a sketch).
These correlations are long-ranged by default as a consequence of the setup, which makes the tMPS representation impractical. 
To overcome this problem, it has been proposed to ``fold'' the network so that both contours are merged via a vectorization operation \cite{banuls2009}: $\bra{\psi(T)}O\ket{\psi(T)}=\bra{\varphi}(\mathbb{I}\otimes O)\ket{\psi(T)}\otimes\ket{\overline{\psi}(T)}$, where $\ket{\overline{\psi}(T)}$ is the complex conjugate of $\ket{\psi(T)}$, and the vector $\ket{\varphi}=\bigotimes_{k=1}^N \left(\sum_{i_k=1}^d \ket{i_k} \otimes \ket{i_k}\right)$ is introduced to reproduce the contraction between the forward and return contours. 
In this manner, the transverse contraction of the folded network can lead to tMPSs with a drastically reduced temporal entanglement, allowing for an advantage compared to standard methods (see \cref{fig:corr_contours}(b)).
%To distinguish folded and unfolded tMPSs will use a double and single ket notation respectively, namely $\langle \bra{L}$ when folded while we keep $\bra{L}$ otherwise.

Aside from providing a novel way to possibly circumvent the entanglement barrier, the idea of a transverse contraction can provide a natural bridge towards the formalism of open quantum systems, and more specifically to the idea of an influence functional.
To see this, let us recall that within the path integral formulation of quantum mechanics, open quantum systems can be studied by constructing the path integral of the system plus environment, and subsequently integrating out the degrees of freedom of the latter. 
The result of this integration is what is called an influence functional (IF), corresponding to a function of the time-trajectories of the system. 

The IF encodes the effects of the environment on the system, and allows evolving the latter's reduced density matrix. 
Being a function of time-dependent coordinates,
the IF can be treated as a vector of the multi-time Hilbert space of the system~\cite{petrat2014multi}, 
and represented using a temporal MPS \cite{lerose2021influence}.
In the context of our TN representation, the equivalence between the two becomes clear if we now observe that in the transverse contraction we effectively trace out the degrees of freedom of a part of the many-body system (i.e.\ the environment), precisely encoding their influence on the rest (the subsystem of interest) at different times through the resulting tMPS.
The folding operation provides another step in this direction, as its overlapping forward and backwards time evolution paths reproduce precisely the Schwinger--Keldysh contour, which is typically encoded in the IF construction.

Having established such a connection, one can leverage the powerful machinery of tensor networks to provide an efficient representation of these functionals encoding the dynamical properties of the system~\cite{lerose2021influence, Giudice_PhysRevLett.128.220401, lerose2023overcoming}.
We will further elaborate on this encoding in the following sections, after introducing the relation of these objects with the process tensors.

\section{The connection between process tensors and influence functionals}\label{sec:3}

% ALT: 
\def\vA{{\bf{A}}}
\def\cA{\mathcal{A}}
\def\ptau{{D_\tau}}
\def\taus{{{\tau_1 \dots \tau_n}}}

Process tensors have originally been introduced as a generalization of channels for operators: citing \cite{chiribella2008,chiribella2009}, they were ``introduced as a tool to optimize quantum circuit over a set of unknown gates for a given task''. 
They can also be connected with the idea of multiple-time states, \cite{aharonov2007,leifer2006,leifer2013} discussed in the context of quantum foundations.
In this section, we will introduce the concept of the {process tensor} following the presentation of \cite{dowling2024,pollock2018, pollock2018a}, and
 relate it to the tensorial objects encountered in \cref{sec:2}.

In the study of the evolution of open quantum systems, a significant challenge is that of generalizing the notion of Markovianity of classical stochastic processes \cite{pollock2018}.
A classical stochastic process is defined by the joint probability distribution of a stochastic variable $X$ that describes the state of the system at several instants of time, $P_c(X_n,t_n ; X_{n-1},t_{n-1}; \cdots; X_1,t_1)$.
Markovianity %is equivalent to the property 
implies that the state of the system at a given time only depends on its state at the immediately previous time, 
\begin{equation}
P_c(X_n,t_n ;  \cdots ; X_1,t_1)%P_c(X_k,t_k; \cdots) 
= P_c(X_n t_n;X_{n-1},t_{n-1} ) \,.
\label{Markov}
\end{equation}

\begin{figure}[t]
  \centering
  \includegraphics[width=0.78\linewidth]{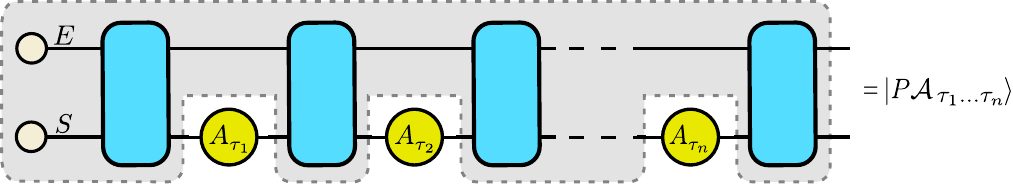}
\caption{\label{fig:process_tensor_qi}%
%\stc{TODO update caption}
The process tensor is represented as the gray shadowed area, containing the information on the initial state of the system (S) and its environment (E), together with their joint evolution under the unitaries $U(t_i,t_{i-1})$. The evolution of the system is monitored by control operations $ \cA_\taus = \set{ A_{\tau_1}, \dots A_{\tau_{n}} }$ performed at times $t_1,\dots t_n$. The process tensor maps the chosen control operations to the output state of the system plus the environment, $\ket{P \cA_\taus}$.
}
\end{figure}

The generalization of these ideas to the quantum realm is not straightforward,
since the measurements needed to determine the control parameters X will disturb the process that we want to characterize. 
The multi-time process tensor formalism overcomes this problem by establishing a clear separation between the process and the control operations performed by the observer, leading to a definition of Markovianity that reduces to \cref{Markov} for classical processes.

The control operations are completely positive trace non-increasing actions on the system, representing for example a unitary transformation (trace preserving) or a possible result from a measurement (trace decreasing).
These operations  are often called {\emph{instruments}}. We might be interested  on a set of available instruments at an instant of time, each of them to be chosen with a given probability. We represent the set of available instruments as Kraus operators $\vA = \set{A_j}$ defining a quantum channel. The resulting quantum channel must also be trace non-increasing  
\begin{equation}
\sum_{j}  A^{\dagger }_{j} A_{j} \leq \mathbb{I} \,.
\label{eq:instruments}
\end{equation}
%can be described by a set of norm-non-increasing operators $\vA = \set{A_j} , j = 1, \dots d^2$,  representing for example the result of a measurement or a unitary transformation. These operations  are often called {\emph{instruments}}. The set of all instruments that can be performed at a given time defines a trace preserving completely positive map \stc{, TODO start from here? though for our intents we can restrict ourselves to a subset satisfying 
%\begin{equation}
%\sum_{j}  A^{\dagger }_{j} A_{j} \leq %\mathbb{I} \,.
%\label{eq:instruments}
%\end{equation}
% }

We consider 
%Let us now consider the action of 
that the control operations intervene at $n$ different instants of time $t_1, \dots t_n$. %(here and in the following we omit for brevity the $T$ subscript). 
The 
\emph{process tensor} $P$ is a linear and completely positive map 
 from the set of control operations
 $ \cA = \set{ \vA^{[t_1]} ,\ldots,  \vA^{[t_{n}]} } $ acting on an open system, to output states of the system plus its environment. The output states can be sub-normalized according to the success probabilities of the chosen instruments.
 The process tensor thus encodes the system and environment dynamics, as well as the information on their initial state. In order to simplify notation and 
 anticipating the discussion below, we assume that the same set of instruments is available at each instant of time, i.e., 
 $\vA^{[t_1]} = \dots = \vA^{[t_{n}]} = \vA$. 
 %At this point, we can begin making a connection with the tensor network formalism.
 We denote a particular choice of instruments by $ \cA_\taus = \set{ A_{\tau_1}, \dots A_{\tau_{n}} }$.
 The process tensor would then map 
 \begin{equation}
P:  \quad \cA_\taus \, \rightarrow \, %P(\cA_\taus) \equiv 
\ket{P\cA_\taus} \in {\mathcal{H}}_{S\otimes E} \,,
\end{equation}
with
 \begin{equation}
\ket{P\cA_\taus}
= U(T,t_n) A_{\tau_{n}} \cdots \,U(t_2,t_1) A_{\tau_1}U(t_1,0)\ket{\psi^0_{SE}} \ ,
\label{eq:PTofA}
\end{equation}
where  $\ket{\psi_{SE}^0}$ is the initial state of system plus environment and $U(t_i,t_{i-1})$ implements their joint unitary evolution between times $t_{i-1}$ and $t_i$.
 \Cref{fig:process_tensor_qi} displays a graphical representation of \cref{eq:PTofA}.
 
%We make now a step further and associate the labels of the instruments with elements of a ``temporal'' Hilbert space, ${\mathcal{H}}_T$. This allows to define a state in the enlarged Hilbert space ${\mathcal{H}}_E\otimes {\mathcal{H}}_S \otimes {\mathcal{H}}_T$, containing information on all possible choices of instruments
% \begin{equation}
%  \ket{P\cA} = \sum_{\taus} \ket{P\cA_\taus} \ket{\taus} \,.
%  \label{PA}
%  \end{equation}

\begin{figure}[t]
  \centering
  \includegraphics[width=1\linewidth]{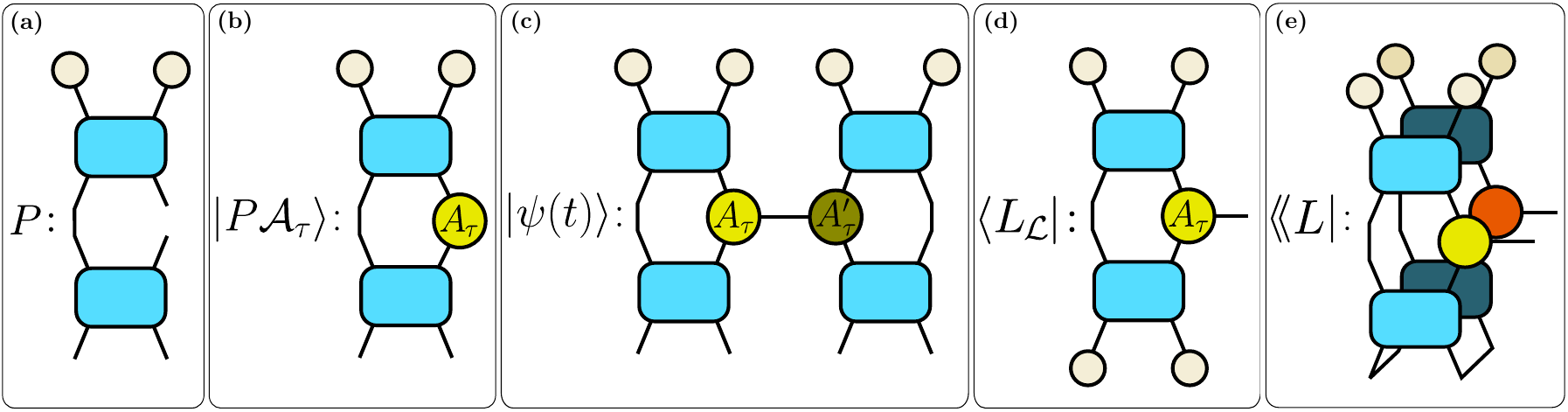}
  \caption{\label{fig:process_tensor}
  %\stc{TODO this caption needs to be updated}
      Tensor network representation of 
      (a) the process tensor $P$ for two spins and two time steps, and (b) the process tensor with an instrument $A_\tau$ applied. %In both cases, the process tensor contains spatial and temporal legs.
      (c) Evolution of a four spin system where, according to \eqref{UAA}, the central cyan gate is interpreted in terms  of instruments applied to the associated left and right process tensors. 
      %When the instrument corresponds to the factorization of a unitary gate, as described in \cref{fig:unitarity}, one has to absorb the singular values (red dot in (c.1) ) and the normalization factors. , necessary to define the instruments as an effective dissipative evolution $p_t$ as for \cref{eq:rel_P_L} represented as a blue half circle in the figure. 
      %(d) The partial trace of the process tensor over its spatial degrees of freedom gives rise to the influence functional of half of a system made by four spins, whose evolved state $ \ket{\psi(t)}$ is represented in (f). Such process tensor is a mixed state $\rho_L$ which can be vectorized as $\bra{\langle L}$. 
      (d) The left temporal vector in the context of a Loschmidt echo represented as the overlap of the spatio-temporal state $|P \cA\rangle_L$ \eqref{PA} on the initial state of the spins. The open leg corresponds to the promotion of the instrument labels to elements of a temporal Hilbert space ${\mathcal{H}}_T$.
      % is the overlap of the process tensor with the part of the initial state with support on $L$, in this case the two leftmost spins. 
      (e) The partial trace of $|P \cA\rangle_L$ over its spatial degrees of freedom gives rise to the influence functional of the two left spins. % made by four spins, whose evolved state $ \ket{\psi(t)}$ is represented in (f).
  }
\end{figure}

Let us proceed to show how 
the language of process tensor naturally emerges in the study
%such an object naturally appears in the definition 
of the closed dynamics of a quantum spin chain. 
Imagine that we start by the simplest scenario obtained by considering the evolution of two constituents.
In order to make connection with  \Cref{fig:process_tensor_qi}, we can assume that one plays the role of the system and the other of the environment. 
We also limit ourselves to the case of a two-times process tensor where an initial product state evolves under the unitary $U$ for one time step, then might undergo a control operation and finally evolves again for an extra step, as shown in \cref{fig:process_tensor}(a).
For this setup, the process tensor has four open legs: two (one ``input'' and one ``output'')  after one step of evolution, which 
on the introduction of an instrument get mapped to the \emph{temporal} legs defined above, and two output legs after the second step, which are \emph{spatial} legs corresponding to the two constituents. 
If we contract the former two legs together (corresponding to a trivial choice of instrument, $A=\mathbb{I}$), the latter will represent the final state of such constituents.

Among the possible choices of instrument sets, there is one which is relevant for the connection we want to make here:
it consists in using as instruments the rows (columns) of the left (right) unitary emerging from the decomposition of the evolution operator shown in \cref{fig:unitarity}(a), $U=U_L \Sigma V_R$. 
Namely, we can choose a set $\vA=\set{A_\tau}$ such that 
\begin{equation}
\big[A_{\tau}\big]_{\alpha \beta} = 
\frac{1}{\sqrt{d}} 
%{\sqrt{\lambda_\tau} \over d}
\big[U_L\big]_{\alpha \beta;\tau} \,,
\label{eq:instr_from_te}
\end{equation}
with $\tau=1,..,d^2$. The normalization $1/\sqrt{d}$ renders the quantum channel defined by $\vA$ trace preserving (see \cref{fig:unitarity}(b)),
\begin{equation}
    \sum_{\tau=1}^d \Big[ A^\dagger_\tau A_\tau \Big]_{\alpha \gamma}  = {1 \over d} \sum_{\tau,\beta} \big[U_L^\ast\big]_{\beta \alpha;\tau} \big[U_L\big]_{\beta \gamma;\tau}= % \frac{1}{d} \sum_\tau 
    \delta_{\alpha \gamma}  \,.
\end{equation}

The set $\vA$ so defined assigns equal probability to each instrument. Although a valid choice, this however does not represent the probabilities with which the $A_\tau$ get selected as the system evolves. That information is encoded in the singular values $\sigma_\tau$, contained in the diagonal matrix $\Sigma$. Since $U$ is unitary, we have $\sum \sigma^2_\tau=d^2$ and hence each singular value satisfies $\sigma_\tau/d\leq 1$. We can use this fact to rescale the instruments as
\begin{equation}
    A_\tau \to \sqrt{\sigma_\tau \over d} A_\tau \ ,
    \label{AU}
\end{equation}
such that now $\vA$ will describe in general a trace decreasing quantum channel. We denote as $A'_\tau$ the alternative set of instruments associated with the right unitary $V_R$.
In this way, the decomposition of the two-body gates can be rewritten as
\begin{equation}
U=d^2 \sum_\tau A_\tau \, A'_\tau \ .
\label{UAA}
\end{equation}

This specific choice of instruments leads us to the first connection between the multi-time process tensor and the influence functionals.
To see this, let us now consider a system of four qudits evolving under a brick wall circuit based on the same two-body gates as before. %\stc{TODO Check refs to Fig5}
Splitting
the system in half,
we consider that the left and right qudits undergo as before two evolution steps, but now the central left and right  qudits are connected by one further step. We can interpret such step as the insertion of the instruments $A_\tau$ and $A'_\tau$ introduced above on a left and right process tensors, defined as in \cref{fig:process_tensor}(a). This is shown in \cref{fig:process_tensor}(d), where a sum over instruments
is understood. We make a step further and associate the labels of the instruments with elements of a ``temporal'' Hilbert space: $\tau \to \ket \tau \in {\mathcal{H}}_T$. This allows us to define  spatio-temporal states in the enlarged Hilbert spaces $ {\mathbb{C}}^{d} \otimes{\mathbb{C}}^{d} \otimes {\mathcal{H}}_T$ and ${\mathbb{C}}^{d} \otimes{\mathbb{C}}^{d}\otimes {\mathcal{H}^\ast_T}$ as follows
\begin{equation}
\ket{P\cA}_R = \sum_{\tau} \ket{P A'_\tau} \ket{\tau} \, , \hspace{1cm}  \ket{P\cA}_L = \sum_{\tau} \ket{P A_\tau} \bra{\tau} \, .
 \label{PA}
\end{equation}
Recall that both the left and right process tensors and the instrument sets $\cA$ and $\cA'$ are constructed out of the same object, the 2-body evolution gate $U$. Hence the same information is contained in the process tensors and the spatio-temporal states \eqref{PA}, particularly when it comes to the computational complexity of encoding these objects, and in this sense we can use them interchangeably. 

Suppose we are interested in the Loschmidt echo, \cref{eq:loschmidt_echo}, of the four spins,  which according to the results of \cref{sec:2} can be computed as the overlap of a left and right temporal vectors $\braket{L_{\mathcal L}| R_{\mathcal L}}$. 
%Splitting the system in half, we can introduce a left and a right process tensors as in \cref{fig:process_tensor}(a), which are connected via the leg shared by the instruments.
Following the discussion above, the left temporal vector is obtained by projecting the space-like degrees of freedom of $\ket{P \cA}_L$
%the spatio-temporal state \eqref{PA}, associated with the left process tensor and the choice of instruments \eqref{AU}, 
onto their initial state,
% Following the discussion above, the wave-function of the left temporal vector is then equivalent to the projection of the space-like degrees of freedom of the left process tensor onto the initial state, up to the aforementioned Euclidean time evolution, 
\begin{align}
\bra{L_{\mathcal{L}}} &= d  \braket{\psi_{0L}|P\cA}_L \label{eq:rel_P_L}  \,.
%_L\braket{\psi_0|P\cA}_L =  \langle{L}_{\mathcal{L}} | \, \ptau^{-1} \ 
\end{align}
%The subscripts $L$ refers to the fact that we are only considering the first two spins on the left. 
We are assuming that the initial state of the four spins is a product state, such that we can assign a well defined initial state
to the left two spins,  $\ket{\psi_{0L}}$.
%$\ket{\psi_0}$ to be a product state, and thus we can assign a well defined initial state to the two left spins, $\ket{\psi_{0L}}$.
This equivalence is represented in \Fig{fig:process_tensor}(c). 
Following an analogous reasoning, we can identify a right process tensor and the corresponding temporal vector as $\ket{R_\mathcal{L}} = d \braket{\psi_{0R}|P\cA}_R$. 

%\stc{=== New }

We turn now to the network
%For our final step towards the influence functional, let us start again with our four-component setup and construct the network 
associated with another typical dynamical quantity of interest, 
namely the expectation value of an operator. Focusing on the left half of our 4-spin system, this network will contain both %the process tensor 
$\ket{P\cA}_L$  %introduced above
 and its conjugate $\bra{P\cA}_L$, which can be associated with the description of a backwards time evolution. 
The trivial case of an identity operator in the system+environment legs immediately leads to a partial trace obtained by multiplying the two process tensors over their open spatial indices, as illustrated in \Fig{fig:process_tensor}(e). 
 Of course, the result can be generalized with the inclusion of different operators, inserted in the spatial legs of the process tensors, leading 
 to the same structure.
The resulting object, describing time evolution on a time contour involving both forward and backwards evolution, is precisely the influence functional of 
the left half of the system, as introduced in \cref{sec:2}.
In the language of process tensors, this left influence functional can thus be seen as the reduced density matrix of %process tensor 
$\ket{P\cA}_L$ traced over its spatial degrees of freedom
\begin{equation}
    \rho_T^L = d^2 \, {\tr_{\mathrm{space}}}\, _L\!\ketbra{P\cA}_L
\ .
\label{PO}
\end{equation}
If we vectorize $\rho_T^L$ by ``folding" the network and joining the temporal legs associated with the forward and backwards evolution (and recovering the Keldysh contour) we can interpret it as a state $\langle\bra{L}$, as  shown in \cref{fig:process_tensor}(e).
%, we can also see it as a state $\langle\bra{L}$, result of the aforementioned partial trace.

%
\begin{figure}[t]
  \centering
  \includegraphics[width=0.9\linewidth]{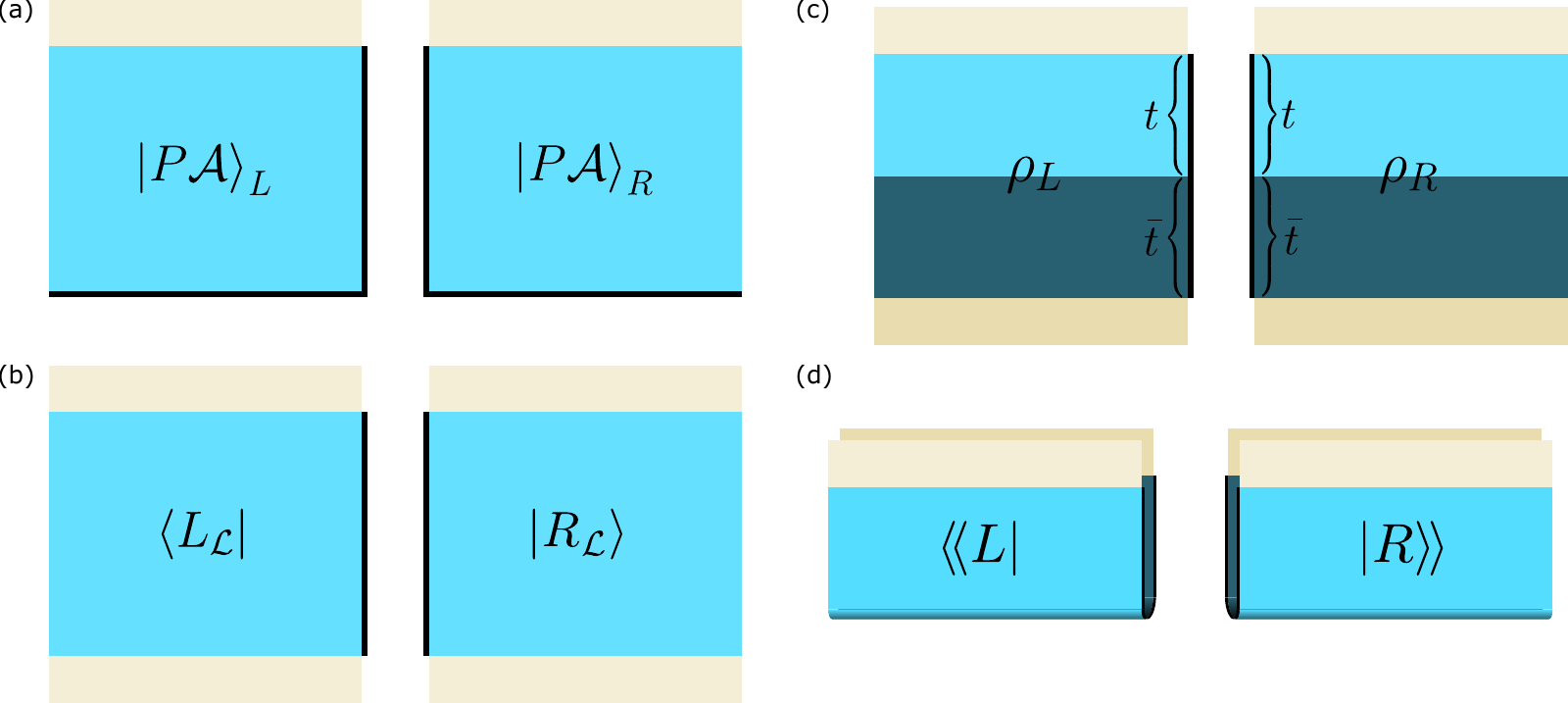}
  \caption{\label{fig:process_tensor_mb}
      \textbf{(a)} 
      Contracting a patch of the spatio-temporal tensor network gives rise to a tensor with both spatial and temporal open legs, represented by black solid lines. When cutting the network, we split the two body gates joining the left and the right halves following the recipe of \cref{fig:unitarity}. The resulting tensor can be identified with the spatio-temporal state $\ket{P \cA}_L$ defined in \eqref{PA}, with the choice of instruments \eqref{AU}. %The same construction holds for the  $\ket{P \cA}_R$ .
      %Contracting a patch of the spatio-temporal tensor network gives rise to a tensor with both spatial and temporal legs, thus to a process tensor. Here we show the specific case in which the patch includes half of the spatial degrees of freedom contracted for the full temporal duration. In this way  the left patch  generates a left process tensor. When cutting the network, we split the two body gates joining the left and the right halves following the recipe of \cref{fig:unitarity}. As a result the process tensors include the instruments defined in \cref{eq:instruments}, $P_L(\set{A})$. The same construction holds for the right process tensor $P_R(\set{A})$. Notice that the black solid lines indicate the open legs of the process tensors. \stc{TODO We don't mention here dissipative $\ptau$ pieces ? } 
      \textbf{(b)} Starting from a product state, we consider the overlap of $\ket{P \cA}_L$ with the initial state of the left half of the system, obtaining the left temporal state defined in the context of Loschmidt echo, $\bra{L_\mathcal{L}}$.
      %Starting from an initial product state, we can consider the overlap of the corresponding process tensor  $P_L(\set{A})$ with the portion of the initial states defined on the left half of the system, obtaining the left temporal state defined in the context of Loschmidt echoes studies $ \bra{{\psi_0}_L} P_L(\set{A}) \equiv \bra{L_\mathcal{L}}$. \stc{TODO update notation in figure matching this} Here the only open legs are indeed along the temporal direction.
      \textbf{(c)} The partial trace of $\ket{P \cA}_L$ on its spatial degrees of freedoms, gives raise to the left influence functional defined in standard tensor networks studies of the dynamics.
      %The partial trace of $P_L(\set{A})$ on its spatial degrees of freedoms, $\tr_{\textrm{space}}[P_L(\set{A})P^{\dagger}_L(\set{A})] \equiv \rho_L$, represents the left influence functionals defined in standard tensor networks studies of the dynamics, which is in general a mixed state.
      \textbf{(d)} We can vectorize and fold the influence functional, obtaining the state $\bra{\langle L}$. The same constructions hold for the right half of the tensor network.
      %As for any density matrix one can also define the vectorized form of  $\rho_L \equiv \bra{\langle L}$ which is usually represented as folded tensor network. Here, we represent both $\bra{\langle L}$ and $\ket{R\rangle}$%
      }
\end{figure}

At this point, it is straightforward to extend these quantities to the many-body setting considered in~\cref{sec:2}. 
We start by considering the tensor network that describes the evolution of the system under $N$ Trotter steps.
We focus on a the partial contraction of such network on a spatio-temporal patch, say eg. the left half of the system. The resulting tensor has spatial and temporal open legs represented by horizontal and vertical black solid lines in
\Fig{fig:process_tensor_mb}(a). We divide its spatial legs in system and environment, such that the system comprises just the rightmost spin of the left patch. In analogy with \eqref{PA}, we can interpret this spatio-temporal tensor in the language of process tensors by defining 
\begin{equation}
 \ket{P\cA}_L = \sum_{\taus} \ket{P\cA_\taus} \bra{\taus} \, ,
\label{PA2}
\end{equation}
where $P$ is a generalization of \cref{fig:process_tensor}(a) and the instruments are derived again from the singular value decomposition of the 2-qubit evolution gate. 
%A partial contraction of such network on a spatio-temporal patch, say eg. the left half of the system, naturally defines a %process tensor state $\ket{P\cA}_L$. This is shown in  \Fig{fig:process_tensor_mb}(a), where horizontal and vertical black solid lines represent respectively open spatial and temporal legs. 
If we now project the spatial legs of $\ket{P\cA}_L$ onto
% the restriction of the initial state for 
the initial state of the chosen spatial patch, 
we obtain the left environment defined in the context of Loschmidt echoes as described in \eqref{eq:rel_P_L} and illustrated in \Fig{fig:process_tensor_mb}(b). %(up to a product Euclidean evolution on the temporal links).
%This is illustrated in \Fig{fig:process_tensor_mb}(b), where we represent both the left and right temporal vectors involved in the calculation of the Loschmidt echo. Relation \cref{eq:rel_P_L} generalizes immediately to  \stc{TODO still to decide on the various $p_i$} 
%\begin{equation}
%\langle{L}_{\mathcal{L}} | = \left( \otimes_{i=0}^{T-1} p_{t_i} \right) _L\braket{\psi_0|P\cA}_L
%\label{eq:rel_P_L2}
%\end{equation}
%where $L$ again refers to the left half of the system (the same can be done for the right half).
%We consider a Trotterized evolution with $T$ time steps.
%   Projecting the temporal legs of the left environment %in Fig. \ref{fig::process_tensors_mb}a
%   onto the state $\ket{\vec t}=\ket{t_0\dots t_{T-1}}$ leads to a state $|\psi_{\vec{t}}(t)\rangle$ on the spatial degrees of freedom
%   This state can be used to define a process tensor for the rightmost spin of the left environment
%   \begin{equation}
%   %   |\psi(t)\rangle=P(\{A_{x}\}) \prod_{i=1}^T p_{x_i}^{-1}\ ,
%   P_L(\{A_{x}\})= \left( \prod_{i=0}^{T-1} p_{x_i}^{-1} \right)|\psi_{\vec{x}}(t)\rangle \ ,
%   \end{equation}
%
%With the instruments $\{A_t\}=\{A_{t_0},\ldots,A_{t_{T-1}}\}$ given by \eqref{eq:instr_from_te}, and $p_{t_i}$ implementing again a necessary correction interpreted as an additional Euclidean evolution of the temporal legs.
% 
%\Cref{fig:process_tensor_mb}(b) represents the left and right temporal vectors involved in the calculation of the Loschmidt echo. 
%Relation \eqref{eq:rel_P_L} generalizes immediately to
% 
% 
In the same way, 
%one can represent in the language of process tensors 
the left influence functional associated with the calculation of local observables is obtained from \eqref{PO} also in the many-body setting. %, as depicted in \Fig{fig:process_tensor_mb} (c)--(d). 
 %provides a graphical representation of \eqref{PO} in the many-body context, 
\Fig{fig:process_tensor_mb}(c) and \ref{fig:process_tensor_mb}(d) display the unfolded and folded versions of the influence functionals, respectively.

%Once more, they are the partial trace of the process tensors sketched in \Fig{fig:process_tensor_mb} (a) over the spatial degrees of freedom,
%\begin{equation}
%    \rho_T^L \equiv \bra{\langle L} = \left( \prod_{\bar{t}=0}^{T-1}\prod_{t=0}^{T-1} p_{{t}}p_{{\bar{t}}} \right) {\Tr_{\mathrm{space}}}\, \ketbra{P\cA}_L
%\,,
%\end{equation}
%where $t$ and $\bar{t}$ label the legs of the temporal forward and return contours respectively.

In the above formulas we have made clear that the influence functional is in general a mixed state. 
If one is interested in the compressibility of such a mixed state in terms of matrix-product operator, one has to consider the operator entanglement of the influence functional, that is, the entanglement entropy of its vectorized form that we have indicated with $\bra{\langle L}$.

In the following section, we will discuss some of the algorithms developed for an efficient compression of the influence functional as temporal MPS
and the complexity of such an encoding.

%With these connections in mind, we can now analyze different types of temporal entanglement. 
%The generalized temporal entanglement discussed in the context of Loschmidt echo, CFT and holography is obtained by building reduced transition matrices (RTMs), defined by the partial contraction of the left and right temporal states, 
%\begin{equation}\label{eq:red_trans_matrix}
%\tau (t) =\frac{\textrm{tr}_{T-t} \ket{R_{\mathcal{L}}}\bra{L_{\mathcal{L}}}}{\braket{L_{\mathcal{L}}| R_{\mathcal{L}}}} \ ,
%\end{equation}
%and computing their Rényi or entanglement entropies. We define those in analogy to their ordinary counterpart as,
%\begin{equation}\label{eq:def-gen-temp-entropies}
%  \mathcal{S}^\alpha(t) = \frac{1}{1 - \alpha}\, \log[ \text{tr}(\tau(t))^\alpha] \,,
%\end{equation}
%with \(0 \leq \alpha \leq \infty\) the Rényi parameter.
%On the other hand, one can compute reduced density matrices (RDMs) of the left or right influence functional, 
%\begin{align}\label{eq:red_density_matrix}
%\rho_L (t) &= \tr_{T-t} \ketbra{L}{L} \ ,\\
%\rho_R (t) &= \tr_{T-t} \ketbra{R}{R} \ ,
%\end{align}
%which correspond to the operator entanglement of the corresponding process tensors. 

\section{Algorithms for obtaining a tensor network encoding of the influence functional, dissipative time evolutions}
\label{sec:4}

The transverse contraction framework introduced in \cref{sec:2} for the description of the dynamics of quantum many-body systems opens the way to an efficient encoding of process tensors and influence functionals in the compact form of temporal MPS.
{The best procedure to contract the network associated with time evolution can vary depending on the system we are consdiering and the truncation procedure we choose, as described in the following.
For a finite chain, one possibility is to start from the sides, identify the left- and rightmost columns as temporal MPS and progressively apply columns of tMPO to them until reaching the center, truncating at each step to prevent their bond dimension to grow exponentially. Alternatively, for an infinite homogeneous system one could start growing the system from the center,
in the spirit of infinite DMRG.}

In this section, we explain the main aspects to take into account for these transverse algorithms, and the compression procedure for these tMPS.
The steps described here can be applied to both pure and mixed states, corresponding to the ``folded'' picture introduced in \cref{sec:2}.
As such, in the following description whenever possible we will consider generic left vectors $\bra{L}$ (same goes for $\ket{R}$), assuming that -- depending on the case considered -- they refer to the Loschmidt echo-type setup  $\bra{L_{\mathcal{L}}}$ or the vectorized density matrices $\langle\bra{L}$ introduced above.

 \begin{figure}[t]
 \centering
 \includegraphics[width=\linewidth]{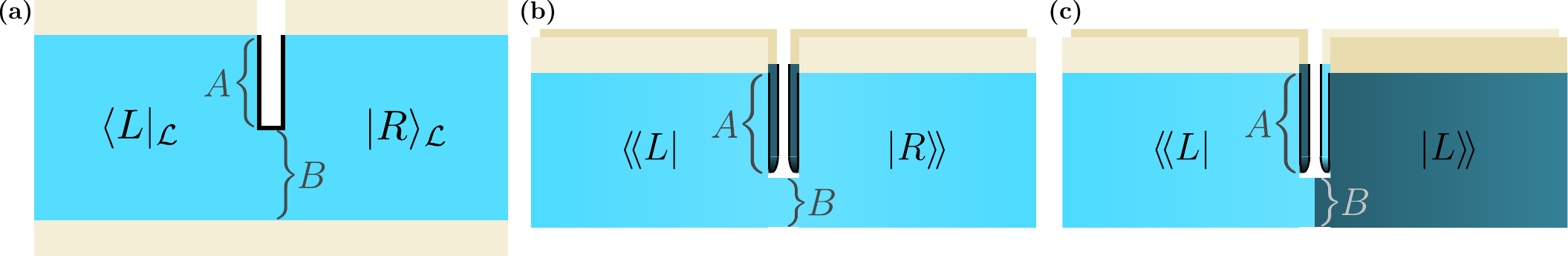}
 \caption{\label{fig:tm}%
   (a) Reduced transition matrices used to compute the generalized temporal entropies from the path integral encoding a Loschmidt echo (b) Reduced transition matrices used to compute the generalized temporal entropy in the context of a quantum quench. Notice that they encode partial overlap of the  the left and right influence functional in that context. 
   Panel (c) shows the  temporal entanglement defined as the entanglement of the vectorized left (or right) influence functional. Notice that, when there is a reflection symmetry c) differs from b) since it requires considering a transposition of the right influence functional before taking the partial overlap. Also given that we are using the vectorized version of the influence functional, its entanglement actually represents the operator entanglement of the influence functional, which is the relevant one if we are interested in understanding how much we can compress the influence functional as an MPO. }
 \end{figure}

\subsection{Cost functions used for compressing tMPS}

Transverse contraction algorithms rely on the successive applications of tMPOs to the tMPS associated with the left and right halves of the system, which we will refer to as  $\bra{L}$ and $\ket{R}$ respectively,
 with a subsequent recompression that can be performed following different criteria.

For this iterative process, %of contracting $\bra{L}$ and $\ket{R}$ with the neighboring columns, 
the simplest prescription would be to compress both $\bra{L}$ and $\ket{R}$ separately, following the standard recipe in MPS literature, through their corresponding reduced density matrices (RDM)~\cite{banuls2009}: for any time $t \in [0, T]$ ($T$ being the total time of the evolution) appearing in the trotterized network, one can define a bipartition in two temporal intervals %$A(t),B(t)$ 
$A=[0,t]$ and $B=[t+\delta t,T]$
and build
\begin{align}
\rho^L_{A}(t) = \tr_{B} \ketbra{L}{L} \ , \hspace{1cm} \rho^L_{B}(t) = \tr_{A} \ketbra{L}{L} \ .
%  \rho^L_{A,B}(t) = \tr_{B,A} \ketbra{L}{L} \,, \quad  \rho^R_{A,B}(t) = \tr_{B,A} \ketbra{R}{R} \,.
  \label{eq:red_density_matrix}
\end{align}
and the same for $\rho^R_{A,B}$.
From these, we can compute the standard entanglement entropies,
\begin{equation}\label{eq:def-gen-temp-entropies}
  \mathcal{S}_\alpha(\rho_A) = \frac{1}{1 - \alpha}\, \log[ \tr_A(\rho_A)^\alpha] \,,
\end{equation}
where \(0 \leq \alpha \leq \infty\) is the Rényi parameter.
As is well known, for this prescription based on RDMs the complexity of constructing $\bra{L}$ and $\ket{R}$ can be related to these $S(\rho_A)$, 
which we refer to as temporal entanglement entropies.

The transverse contraction framework offers however more possibilities when it comes to the optimization of the left and right vectors. 
The first insight in this direction was presented in \cite{hastings2015}, where 
it was realized that the imaginary time evolution equations for the the RDM of the right vector\footnote{As usual, the same reasoning can also be applied also to $\bra{L}$} can be rephrased as a DMRG algorithm for a  system mirrored across the bipartition, with a slightly modified Hamiltonian. This can be intuitively seen by relating the (temporal) RDM $\ketbra{R}{R}$ to the evolution of a system made of two copies of the right subsystem, where one follows the same equations as the other, up to a complex conjugation.  
In turn, this inspired the introduction of a modified cost function, built from the right vector and its transpose: $\tilde\rho_R \propto \ketbra{R}{\bar{R}}$ (where the complex conjugation is performed to "undo" the conjugation of the bra). An optimization with respect to this object was shown to provide a good estimate for the whole left-right contraction of the network.
{As a matter of fact, for a system with reflection symmetry, 
$\bra{\bar{R}}$ coincides with the left vector $\bra{L}$. 
This observation then motivates  us to focus directly on optimizing the overlap  $\braket{L|R}$, by providing a faithful } representation of the reduced {\emph{transition matrices}} (RTM) \cite{Carignano_PhysRevResearch.6.033021}, which for an arbitrary bipartition $A-B$ read
\begin{equation}\label{eq:red_trans_matrix}
\tau_{A}(t) = \tr_{B}  \frac{ \ket{R}\bra{L}}{\braket{L|R}}  \, ,
\end{equation}
and analogously for $\tau_B(t)$.
Following this intuition, the complexity of performing the network contraction should be related to properties of the RTM \eqref{eq:red_trans_matrix}, rather than the RDMs defined in \cref{eq:red_density_matrix}. 
{In \cref{fig:tm} we show a graphical representation of these quantities, comparing the RTMs built for the Loschmidt echo and the (folded) IF with the regular RDM.  }

An extended notion of entropy based upon the RTMs can be defined by substituting $\rho_{L,R}(t)$ by $\tau(t)$ in \eqref{eq:def-gen-temp-entropies}, resulting in the so called {\emph{generalized}} entanglement and Rényi entropies \cite{nakata2021,doi2023a}. 
The relevant quantities for the  evaluation of such generalized entropies are the eigenvalues of the RTM, which are not necessarily positive, or even real. Consequently, the generalized entropies are complex-valued quantities. 

The relation of these interesting new quantities with the complexity of representing the associated tensor network is still the object of investigation~\cite{hastings2015,Carignano_PhysRevResearch.6.033021,carignano2024}. In fact, 
the compression procedure based upon RTMs is not straightforward: unlike the (reduced) DMs, $\tau(t)$ is a non-hermitian matrix. As such, eigenvalues and singular values do not coincide in general, and it is not a priori clear which of them use as guideline for the network compression. 
In the following, we will use the singular values as cost function for truncation, as they provide a well-defined set of positive and real quantities.  

It is also relevant to discuss the role of gauge freedom in these non-Hermitian problems. 
In principle, given a transfer matrix $E(T)$, one can perform a gauge transformation $E(T)\rightarrow \mathcal{X} E(T) \mathcal{X}^{-1}$ such that the 2D tensor network associated with the time evolution remains invariant. 
It has been observed \cite{tang2023matrix} that the temporal entanglement in the $\bra{L}$ and $\ket{R}$ can be arbitrarily reduced by these transformations.
This comes however at the cost of accuracy, since these transformations can shift the relevant contributions to the overlap $\braket{L|R}$ to the tails of singular values,
which get truncated in the required compression procedure. 
This issue %can be circumvented by 
is mitigated when 
considering reduced transition matrices, as these gauge transformations can cancel in computing these objects.
%
%This sensitivity to the gauge freedom is partially removed in case of using a RTM instead: the local gauge transformations $\mathcal{X}=X^{\otimes n_T}$ induces to it a similarity transformation, which does not affect its eigenvalues \cite{tang2023matrix}. \stc{?}
%Indeed, this type of gauge transformation is the one that in principle concerns us for time evolution since the freedom in our setup relies on how we write each tensor of the MPO $U(\delta t)$. That is, it affects every row of the two-dimensional TN independently. 
%Therefore, using the RTM for the compression can be seen as a more robust criterion. 
As such, we will focus on this type of truncation here, while recalling that other methods based on bi-orthogonalization techniques  can also be used when dealing with non-Hermitian transfer matrices \cite{wang1997transfer, xiang1998thermodynamics, zhong2024density}.

\subsection{Compression procedure}

The compression of the tMPS is carried out by constructing the desired cost function, such as the RDM~\cite{banuls2009} or the RTM~\cite{hastings2015, Carignano_PhysRevResearch.6.033021}, and approximating it to a given rank for every possible bipartition.
This is commonly achieved through singular value decompositions of the relevant objects.%, making use of the gauge freedom to access the singular value spectra without the cost of decomposing exponentially large matrices.

\begin{figure}[t]
    \centering
    \includegraphics[%trim={0 9cm 0 0},
    width=0.7\linewidth]{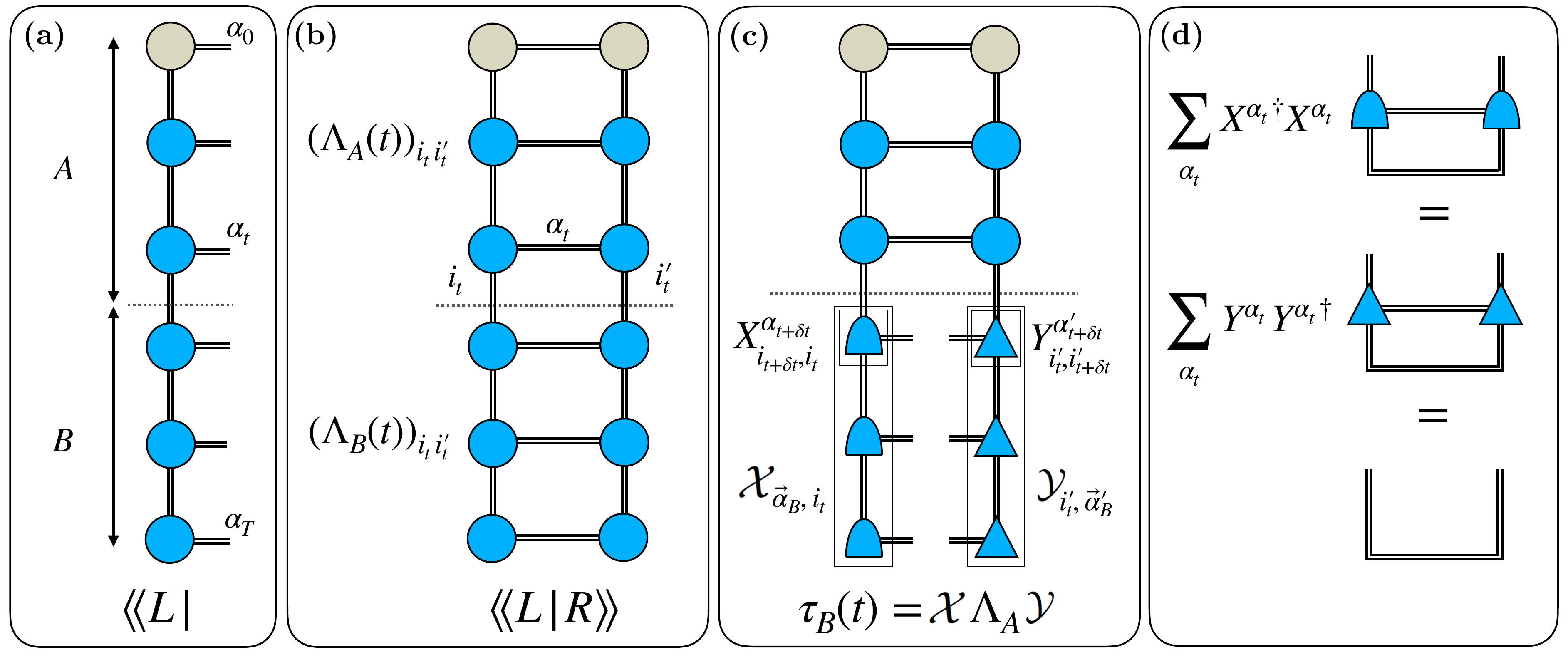}
    \caption{\label{fig:notation_tMPS}%
   Pictorial notation for the compression procedure assuming folded tMPSs. \textbf{(a)} Bipartition of a tMPS into top and bottom subsystems. \textbf{(b)} Definition of $\Lambda_{A,B}(t)$. \textbf{(c)} Reduced transition matrix $\tau_B(t)$ once the bottom subsystems of $\bra{L}$ and $\ket{R}$ are brought into canonical form, i.e., they are expressed in terms of left-normalized $\{X^{\alpha_k}\}$ and right-normalized $\{Y^{\alpha_k}\}$ matrices, respectively. \textbf{(d)} By definition, the matrices $\{X^{\alpha_k}\}$ and $\{Y^{\alpha_k}\}$ satisfy the condition $\sum_{\alpha_k} X^{\alpha_k\,\dagger} X^{\alpha_k}=\sum_{\alpha_k} Y^{\alpha_k} Y^{\alpha_k\,\dagger}=\mathbb{I}$, from which it follows that $\mathcal{X}^\dagger \mathcal{X}=\mathcal{Y}\mathcal{Y}^\dagger=\mathbb{I}$.}
\end{figure}

While the truncation over RDM is amply covered in standard TN literature, let us briefly review the procedure for the case of an RTM.
The notation used is as follows (see \Fig{fig:notation_tMPS}(a)--(b)): given a tMPS, we will use Greek letters for the physical indices, while Latin letters for the bond ones. In addition, given the bipartition introduced above, we define the matrices $\Lambda_A(t)$, which results from the contraction of the subsystem $A$ of $\bra{L}$ with the one of $\ket{R}$, and $\Lambda_B(t)$ as the counterpart for the subsystem $B$.
Notice that $\Lambda_A(t)$ and $\Lambda_B(t)$ can be seen as acting at the bond level, namely $\Lambda_{A,B}(t)\equiv (\Lambda_{A,B}(t))_{i_t\,i'_t}$.

With these definitions, we now proceed to explain the compression procedure. This is done by truncating the RTMs associated with the different bipartitions, which are consecutively obtained by tracing the temporal sites from the $t=0$ to $t=T$, or the other way around. For the sake of concreteness, let us assume the former possibility, although we discuss the relevance of this choice at the end of the section. 
%With these definitions, we proceed to explain the compression procedure. For the sake of concreteness, let us assume that the truncation is done from bottom to top for the RTM \cref{eq:red_trans_matrix}, although, as we will comment at the end of the subsection.
Thus, we start by cutting the tMPS at the first bond index from $t=0$ and bringing the $B$ subsystems of both $\bra{L}$ and $\ket{R}$ into canonical form, namely writing them in terms of left-normalized $\{X^{\alpha_k}\}$ and right-normalized $\{Y^{\alpha_k}\}$ matrices respectively, as depicted in \Fig{fig:notation_tMPS} (c)--(d). 
Once this is done, $\tau_{B}(t)$ reads 
\begin{equation}
  \tau_{B}(t)=\mathcal{X}\Lambda_A(t)\mathcal{Y}\,,
\end{equation}
where $\mathcal{X}$ ($\mathcal{Y}$) represents the left (right) isometry coming from contracting the $\{X^{\alpha_k}\}$ ($\{Y^{\alpha_k}\}$) matrices (see \Fig{fig:notation_tMPS}(c)).
% ,
% \begin{align}
%         X_{(\alpha_{T},...,\alpha_{t+\delta t}), i_t}=\sum_{i_{T-\delta t},...,i_{t+\delta t}} A^{\alpha_T}_{1,i_{T-\delta t}} \cdots A^{\alpha_{t+\delta t}}_{i_{t+\delta t},i_t}\,,
% \end{align}
% \begin{align}
%         Y_{i'_t,(\alpha'_{T},...,\alpha'_{t+\delta t})}=\sum_{i'_{t+\delta t},...,i'_{T-\delta t}} B^{\alpha'_{t+\delta t}}_{i'_{t},i'_{t+\delta t}} \cdots B^{\alpha'_T}_{i'_{T-\delta t},1}  \,.
% \end{align}

Given that $\mathcal{X}$ and $\mathcal{Y}$ are left and right isometries, respectively, $\tau_B(t)$ and $\Lambda_A(t)$ have the same singular values, so that by means of the gauge freedom of the tMPSs the optimization problem has been reduced to applying an SVD to $\Lambda_A(t)$ and discard the smallest singular values, instead of decomposing the exponentially large matrix $\tau_B(t)$. The compressed tMPSs are obtained by inserting the associated projectors on the link which specifies the bipartition. 
Furthermore, a subsequent gauge transformation can be performed on the updated tensors to guarantee that $\Lambda_A(t)$ reduces to an identity.
Once optimized the first link, we only have to update $\Lambda_t$ by tracing out the adjacent site
\begin{equation}
  \Lambda_A(t+\delta t)=\sum_{\alpha_t} X^{\alpha_t} \Lambda_A(t) Y^{\alpha_t}\,,
\end{equation}
and repeat the process until all the bonds are truncated.

\begin{figure}[t]
    \centering
    \includegraphics[width=0.7\linewidth]{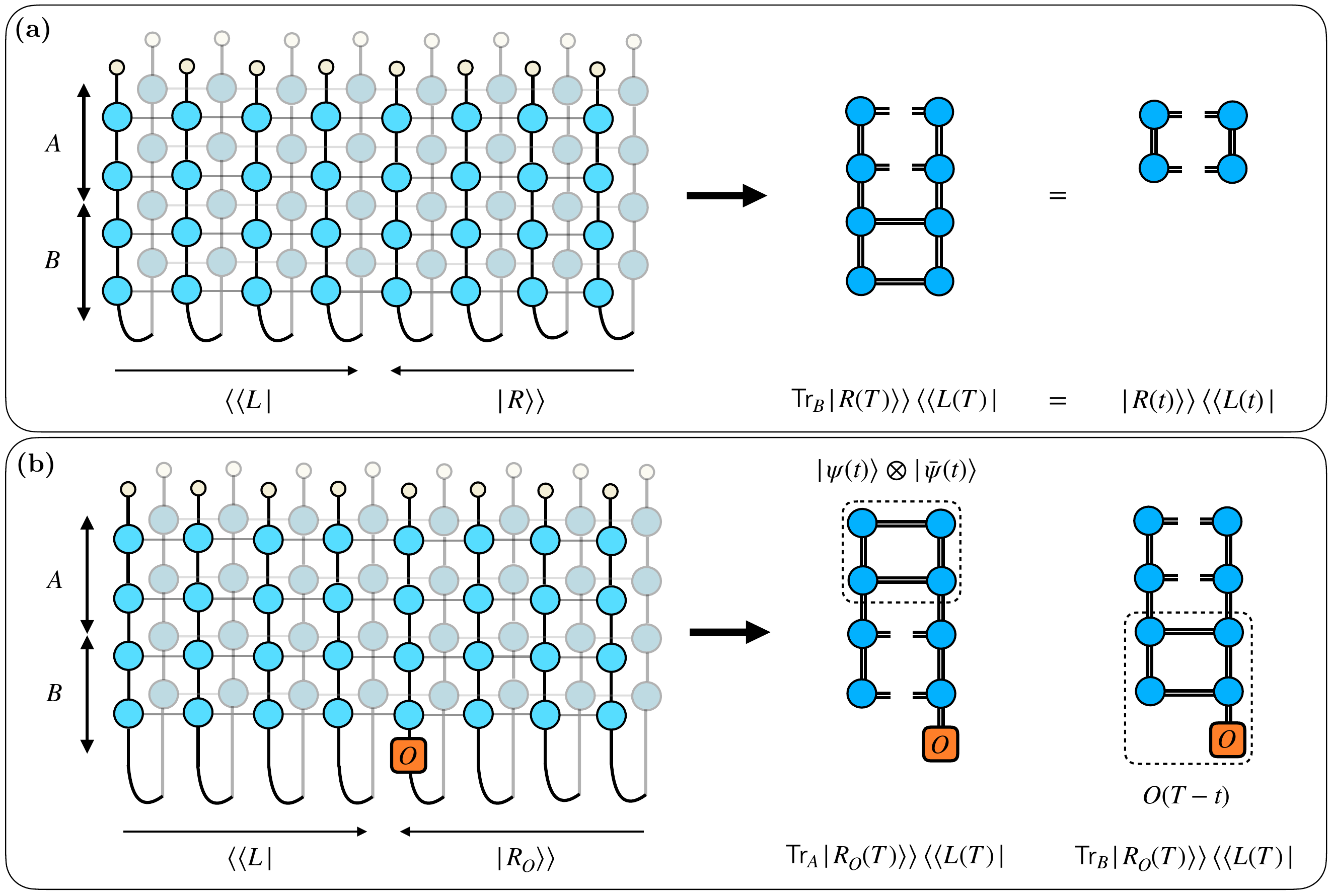}
    \caption{\label{fig:RTMs}%
    Two-dimensional TN representing the RTMs \cref{eq:red_trans_matrix} and \cref{eq:red_trans_matrix_op}. \textbf{(a)} In case of not including the operator, the partial contraction from $t=T$ gives $U^\dagger(T-t)U(T-t)=\mathbb{I}$, so that the RTM $\text{Tr}_{B} \ket{R(T)}\rangle \langle\bra{L(T)}$ equals the projector $\ket{R(t)}\rangle \langle\bra{L(t)}$. Because of this simplification, the associated $\Lambda_B(t)$ has only one non-zero singular value, hence the compressed tMPSs have bond dimension 1 when starting from the bottom side. \textbf{(b)} If the operator is added, then there is a clear connection with the evolution of either the state or operator depending on whether the partial contraction is done from the top or bottom, respectively.}
\end{figure}

It should be noticed that, depending on which dynamical quantity we are interested in as well as the cost function used, a meaningful truncation can require starting from $t=0$ or $t=T$. 
This is especially the case if one compresses the left and right tMPSs for influence functionals through their RTM \cref{eq:red_trans_matrix} in the folded network.
As an example, if one follows Ref. \cite{hastings2015}, the compression must be carried out from $t=0$ to $t=T$: as the RTM constitutes a 2D TN which does not include any local operator $O$, the contraction of $B(t)$ represents the multiplication of the time evolution operator with its adjoint, what gives the identity (see \Fig{fig:RTMs} (a)). Due to this trivialization, if one performs a truncation starting from $t=T$, the states $\bra{L}$ and $\ket{R}$ after the compression would have bond dimension $1$, and contain no useful information for further calculations. This simplification when starting from $t=T$ can be avoided by including $O$ in the definition of the RTM, as proposed in Ref. \cite{Carignano_PhysRevResearch.6.033021}. This is done by contracting the associated transfer matrix $E_O$ with either $\langle \bra{L}$ or $\ket{R}\rangle$, such that the resulting tMPSs are denoted by $\langle\bra{L_O}$ and $\ket{R_O}\rangle$ respectively. Assuming that the operator $O$ is included in $\ket{R}\rangle$, we can define the RTM
\begin{align}
\label{eq:red_trans_matrix_op}
    \tau^{O}_{A} (t) = \frac{\text{tr}_{B} \ket{R_O}\rangle\langle\bra{L}}{\langle \braket{L|R_O} \rangle}\, .
\end{align}
%\stc{although to avoid ambiguity, we will assume in the following that the operator $O$ is included in $\ket{R}\rangle$.}

By following this prescription the contraction of the bottom rows gives the Heisenberg evolution $O(T-t)=U^\dagger(T-t) O U(T-t)$ instead, which does not simplify (see \Fig{fig:RTMs} (b)). In fact, as we explain in \cref{sec:5} A, the inclusion of the operator in the RTM allows for establishing a neat picture on the complexity of computing time-evolved expectation values, based on the concept of the operator entanglement.

This extra care concerning the directionality of the compression is a consequence of having both forward and return contours merged through the folding operation. For this reason, quantities with only the forward contour such as the Loschmidt echo do not present this problem. Indeed, for the Loschmidt echo starting from either the $t=0$ or $t=T$ can be seen as compressing from the side of a state evolving forwards or backward in time, respectively.%, so both possibilities are completely equivalent. %\stc{(Maybe this last sentence induces the question of whether starting from the top or from the bottom give different results in the case of expectation values)}

\section{On the complexity of process tensors and of the influence functional}\label{sec:5}

Having elucidated the connection between temporal MPS, influence functionals and process tensors, we can now employ the techniques described in the previous section in order to complete the picture on the cost of encoding influence functionals as tensor networks, filling some of the gaps in the current understanding of their complexity.

We begin by recalling some results on tMPS complexity in the literature, then discuss their implications for process tensors.

\subsection{Tensor network results based on generalized temporal entanglement}

As already discussed in \cref{sec:4}, in the framework of transverse contractions the generalized temporal entropy can be seen as a measure 
describing the complexity of a faithful description of the network contraction using tMPS.

\subsubsection{Expectation values of local operators} 
A first result on the complexity of the tMPS associated with the time evolution of the expectation value of a local operator was given in Ref.~\cite{Carignano_PhysRevResearch.6.033021}. As explained in \cref{sec:4} B, one can define folded tMPSs with the insertion of the operator $O$ so that the complexity of calculating $\braket{O(T)}$ is given in terms of the ranks of the RTMs \cref{eq:red_trans_matrix_op}. Depending on whether the RTM is obtained by tracing the top or bottom subsystems, we explicitly construct the vectorized time-evolved density matrix of the initial state, $\ket{\psi(t)}\otimes\ket{\overline{\psi}(t)}\equiv \Lambda_A(t)$, or the evolved operator $O(T-t)\equiv \Lambda^O_B(t)$ respectively (see \Fig{fig:RTMs} (b)), fact that can be used to determine the behavior of the ranks with time.

This is especially true for $\tau^{O}_A(t)$, which can be expressed as the multiplication of $\Lambda^O_B(t)$ with the $A$-subsystems of $\langle \bra{L}$ and $\ket{R_O}\rangle$, as illustrated in \Fig{fig:RTMs}(b). The $A$-subsystems of the vectorized influence functionals can be described in terms of $({\bar\Lambda}^{L,R}_A(t))^{1/2}$, where ${\bar\Lambda}^{L,R}_A(t)$ are the analogous of $\Lambda_A(t)$ for the networks $\langle \braket{L|L}\rangle$ and $\langle \braket{R_O|R_O}\rangle$ respectively, see \cite{Carignano_PhysRevResearch.6.033021} for details. Notice that the matrices ${\bar\Lambda}^{L,R}_A(t)$ carry only information on the evolution of the state. Indeed, if we further assume reflection-invariant systems, we have ${\bar\Lambda}^{L}_A(t)={\bar\Lambda}^{R}_A(t)$.
We thus have
\begin{equation}
\rank\left(\tau^{O}_A(t)\right) \le \min \left\{\rank\Big(\Lambda^O_B(t)\Big),\rank\Big({\bar\Lambda}^{L}_A(t)\Big)\right\} \,,
 \label{eq:central_result}
\end{equation}
where $\mathcal{R}$ stands for the rank of the corresponding matrix.
Since the rank of $\Lambda^O_B(t)$ is dictated by its operator entanglement, we obtain that the latter
%Thus, the operator entanglement contained in $\Lambda^O_B(t)$ 
provides an upper bound to the computational complexity of the transition matrices. If its growth is slower (i.e., sub-volume law) than that of the entanglement of the time-evolved state, this implies that the transverse contraction can provide a more efficient compression of the TN associated with the expectation value, a property which can be exploited numerically by truncating the tMPS starting from the operator side \cite{Carignano_PhysRevResearch.6.033021}. 
Explicitly, if $\rank \left(\Lambda^O_B(t)\right)$ only increases polynomially with $T$, we have
\begin{equation}
 \rank \Big( \Lambda^O_B(t) \Big) \le t^{\alpha} \, \Rightarrow \, \rank \left( \tau^{O}_A (t) \right) \le T^{\alpha} \quad \forall \, t \,.
 %\in \set{0,T}\,,
\label{eq:main_bound}
\end{equation}

On the other hand, in the case of ergodic dynamics one expects that the operator entanglement grows linearly with time~\cite{prosen2007operator}, implying 
an exponential cost in representing faithfully the IF for this problem. 
This was confirmed in \cite{Carignano_PhysRevResearch.6.033021}, where some of the specific cases considered were found to saturate this bound.

\subsubsection{Loschmidt echo} 
The transverse contraction framework also allows to provide
an analytical estimate of the computational complexity of evaluating the TN associated with the Loschmidt echo $\braket{\psi_0|\psi(T)}$ 
for an infinite system after a global quench at a critical point, as shown in~\cite{carignano2024}. 
This result was obtained
 by exploiting the universal properties of the model at criticality, and deriving the corresponding properties from the underlying conformal field theory (CFT).
The relevant object for a translation-invariant system is the (non-hermitian) transfer matrix $E^{\mathcal{L}}(T)$ (cfr. \cref{sec:2}) associated with spatial translations in the system\footnote{See also \cite{sirker2005,andraschko2014} for related works in this direction}, and its dominant 
left and right eigenvectors, which can be represented as tMPS. 

\begin{figure}[t]
    \centering
    \includegraphics[%trim={0 9cm 0 0},
    width=0.4\linewidth]{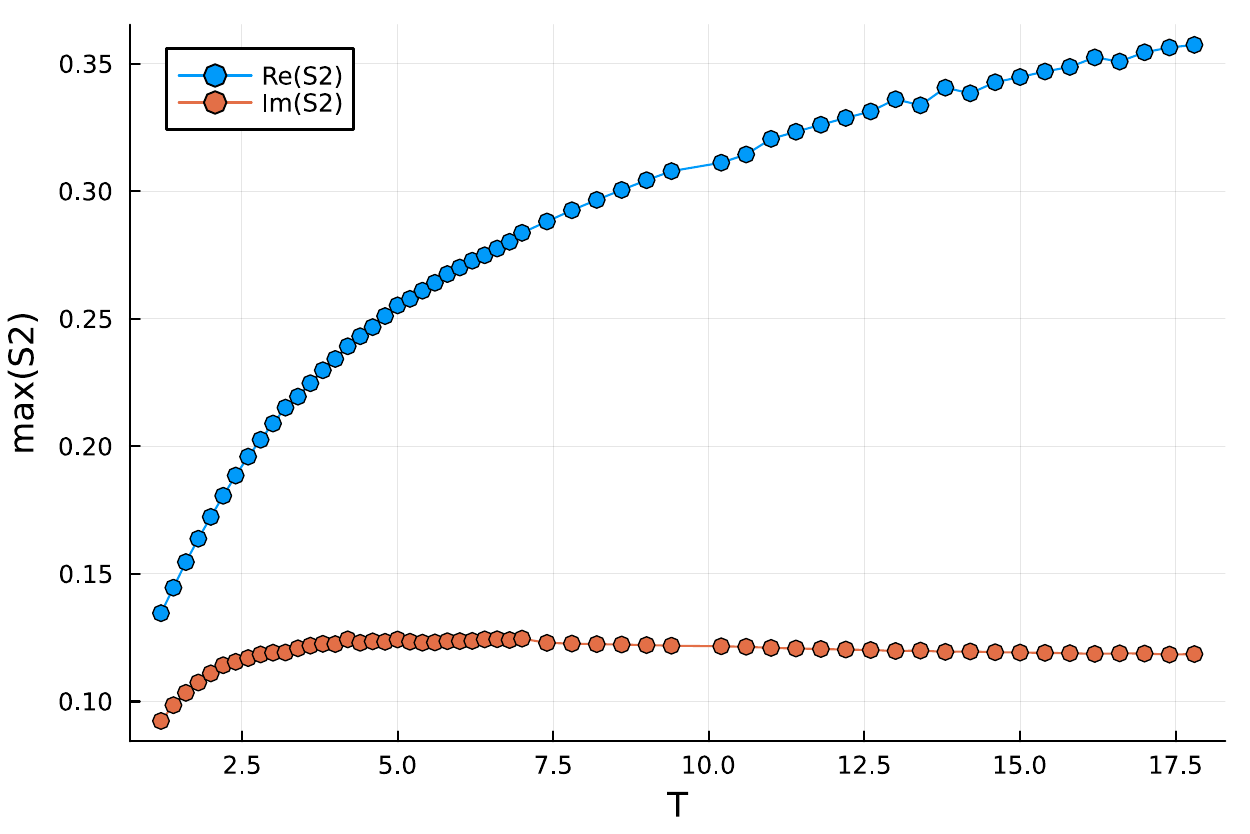}
       \includegraphics[%trim={0 9cm 0 0},
    width=0.4\linewidth]{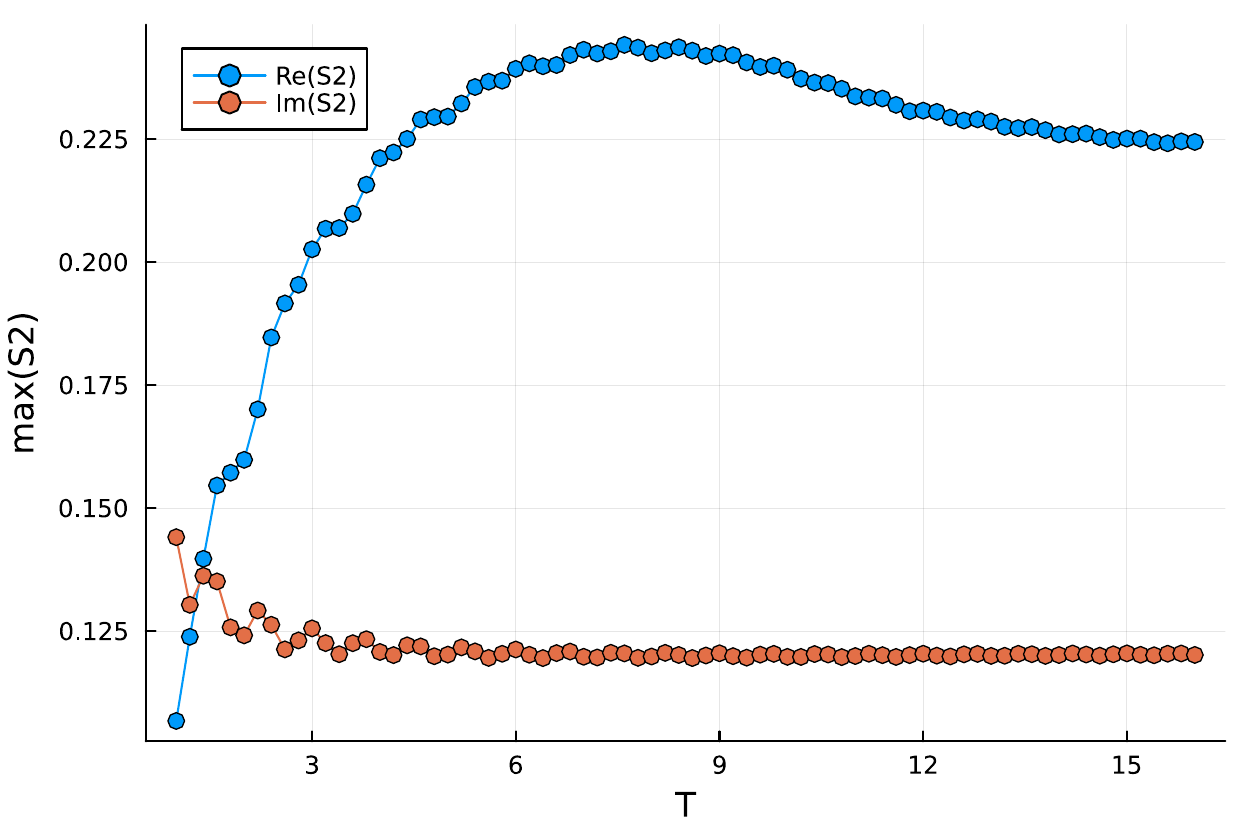}
    \caption{\label{fig:renyi_losch}%
   Examples of generalized second Renyi temporal entropies ($S2$) built from the reduced transition matrix for the Loschmidt echo for the Ising model in a transverse field. We consider a spatially infinite system and plot the maximum of both real and imaginary part of $S2$ as function of time. Left: results at the critical point, where we observe a logarithmic growth. Right: away from the critical point, the entropies saturate exhibiting an area law behavior. }
\end{figure}

The connection in the continuum limit was made again thanks to the path integral formulation, which maps the quench geometry to that of an infinite strip.
CFT can then provide all the relevant information on the spectrum of the transfer matrix along the strip, as well as the generalized temporal entropy associated with a time-like cut, which can be mapped back to the reduced transition matrices $\tau(t)$. 
In particular, it was shown that for a quench to the critical point the generalized temporal entropy grows like $S \sim c \log(T)$, with $c$ the CFT central charge.
This logarithmic growth at criticality (and thus the simulability of the corresponding transition matrix) and the related closing of the gap in the spectrum bear a strong resemblance to what happens in ground states of critical chains and the corresponding  spatial entanglement.
In those cases, the opening of the gap as we move away from the critical point should imply that away from criticality the generalized temporal entropies for the Loschmidt echo follow an area law, guaranteeing an even more efficient simulability using tMPS.
This is indeed the case, as confirmed by numerical simulations in \cite{carignano2024}.  Complementing those results, we plot in \cref{fig:renyi_losch} the maximum value of the Renyi 2 entanglement entropies for the Loschmidt echo at the critical point and close to it, showing the area and logarithmic growths, respectively.

\subsection{Connections with Other Approaches}

It should now be clear that one of the reasons the tensor network community has considered influence functionals and their complexity—measured in terms of temporal entropy—is the hope that these objects might be easier to encode than the evolution of states and operators, which are generally exponentially hard to compress as simple tensor networks.

Recently, we have proposed the use of generalized temporal entanglement, computed from the path integral of the expectation value of local operators, as a measure of the cost of simulating these expectation values \cite{Carignano_PhysRevResearch.6.033021}. Prior studies have instead analyzed the entanglement of independent influence functionals $\bra{L}$ and $\ket{R}$ or the process tensors.

Numerical observations \cite{banuls2009,frias2022} have shown that the temporal entanglement of the influence functional $\bra{L}$ grows logarithmically in some models and linearly in time for others. Subsequent studies \cite{sonner_influence_2021,lerose2021,lerose2021a,Giudice_PhysRevLett.128.220401,lerose2023overcoming} confirmed these findings, associating logarithmic growth with integrability and linear growth with generic dynamics. In \cite{hastings2015}, a direct connection was established between the spatial entanglement of a suitably modified model and temporal entanglement, showing that linear growth in time arises from a non-Hermitian link in the modified model's dynamics, where energy is continually injected. These results were further validated through explicit calculations in random and dual-unitary circuits \cite{foligno2023, Yao2024}, where linear growth was found to be the generic behavior of temporal entropy for the influence functional.

Some authors have suggested that space-time duality might help overcome the entanglement barrier \cite{lerose2023overcoming}. 
The evidence gathered here however points to the contrary: for ergodic systems, influence functionals seem to be exponentially hard to compress using temporal matrix product states (MPS). This may seem counterintuitive, as ergodic systems observed locally are expected to exhibit Markovian dynamics with little or no memory effects. Consequently, one might expect that the quantum channel governing the evolution of a local operator could be encoded using a temporal MPS with small bond dimension, naturally leading to exponential decay of correlations in time. However, we will review why this is not the case.

The first piece of evidence comes from \cite{Carignano_PhysRevResearch.6.033021}, where it was shown that the bond dimension of a temporal MPS encoding the evolution of a local operator is upper-bounded by the bond dimension required to describe the operator's evolution itself. Given the results in \cite{prosen2007operator} and subsequent works, the bond dimension necessary to describe operator evolution in the Heisenberg picture is expected to grow exponentially in time for ergodic systems. Such a growth has been even proposed as a definition of ergodicity. Consequently, the bond dimension of the temporal MPS encoding the influence functional is also bounded by an exponential function of time. 

Building upon these insights, we incorporate findings from quantum information studies on the complexity of the process tensor in ergodic systems. Notably, the definitions of ergodicity and quantum chaos remain debated. Here, we adopt the operational definition introduced in \cite{dowling2024}, which generalizes the classical notion that chaotic systems exhibit extreme sensitivity to perturbations. In this framework, they define "orthogonality of butterfly flutters," meaning that two orthogonal choices of instruments applied to the same process tensor should produce orthogonal system-environment states. Furthermore, such orthogonality should not be rectifiable by a finite-depth unitary transformation, encapsulating the notion of information scrambling. The authors of \cite{dowling2024} demonstrate that, under this definition, the process tensor has maximal entanglement for any space-time cut, implying that its temporal MPS representation must have exponentially large bond dimension.

By taking the partial trace of the process tensor over its spatial degrees of freedom, we deduce that the influence functional—representing a maximally mixed state—is similarly encoded by an MPO with exponentially large bond dimension. Additionally, the fully scrambled entanglement implies that this maximally mixed state cannot be factorized into independent mixed states at each time step, unlike in toy models built from swap tensors \cite{banuls2009,muller-hermes2012}. Furthermore, \cite{dowling2024} establishes that the butterfly flutter definition of ergodicity encompasses the linear growth of local operator entanglement as a special case. Thus, we complement the upper bound obtained in \cite{Carignano_PhysRevResearch.6.033021}, which relies on local operator entanglement growth, with results on the structure of spatio-temporal entanglement in the process tensor, leading to the conclusion that, in general, influence functionals of ergodic systems exhibit linearly growing temporal entanglement and are consequently hard to compress as MPOs.

Parallel findings in the many-body physics community further support these conclusions. Studies on matrix elements of Floquet circuits between initial and final singlet states \cite{ippoliti2021,lu2021,ippoliti2022}—which directly relate to $\bra{L}_{\cal L}$ in the context of the Loschmidt echo—have shown that when system dynamics are ergodic (as defined by the linear increase of entanglement entropy in the evolved state), the temporal entanglement of $\bra{L}_{\cal L}$ also grows linearly with the temporal bipartition size. This confirms that even projections of the process tensor are exponentially hard to compress. This behavior is a converse manifestation of the butterfly flutter phenomenon: in an information-scrambling system, a specific product state has support on exponentially many orthogonal instruments.

A significant implication of these results is that, despite transverse evolution being dissipative (as discussed in earlier sections), its strength in an ergodic system is insufficient to induce an entanglement transition in its stationary states \cite{li2019}, which always remain in a volume-law phase as if the evolution were unitary.

These findings suggest that encoding separate left and right influence functionals in ergodic systems is exponentially hard. However, in other contexts—such as many-body localized dynamics—transverse evolution can induce an entanglement transition, potentially simplifying process tensor encoding \cite{ippoliti2021,lu2021,ippoliti2022}.

Finally, these results raise concerns about the scalability of recent algorithms in open quantum systems. Many assume an environment initially in equilibrium with a thermal bath, which for integrable environments has been shown to yield highly compressible influence functionals \cite{makri1995,strathearn2017,cygorek2022}. However, it remains unclear whether this simplification arises from thermal equilibrium assumptions or from the non-ergodic nature of the environment, which typically consists of free oscillators. A similar question was explored in \cite{ye2021}, where structured baths undergoing unitary dynamics were considered, demonstrating that the compressibility of the environment's influence functional strongly depends on interaction strength. These observations reinforce the idea that strongly interacting, ergodic systems possess complex influence functionals that are inherently difficult to compress.

\section{Conclusions}\label{sec:conclusions}

In these notes, we have reviewed the connections among various objects studied in different quantum physics communities within the unifying framework of tensor networks.
We have reviewed  how both the return probabilities defining the Loschmidt echoes and the time evolution of local expectation values in a closed quantum many-body system can be encoded in the contraction of spatio-temporal tensor networks.
Depending on the type of evolution and the system under consideration, such TNs may involve discretizing the path integral of a continuous system in space and time or simply contracting the quantum circuits that describe Floquet dynamics in lattice models.
Regardless of these specific details, we have shown that contracting spatio-temporal patches of the tensor network results in the same objects: process tensors. These process tensors have been studied for different reasons in the quantum information community.
In the context of tensor networks for characterizing out-of-equilibrium dynamics in many-body quantum systems, these objects are of interest due to the  assumption that they might be easier to compress as matrix product states than the full evolving states.

%\stc{TODO cite latest Loschmidt echo measurement? } 

This view has so far been confirmed only for integrable systems, where we have shown that the complexity of the influence functional is upper-bounded by the local operator entanglement, which is conjectured to grow only logarithmically with time \cite{carignano2024}.
Unfortunately, the same upper bound indicates that for generic ergodic dynamics, the complexity of the influence functional could grow exponentially with time.
By exploiting the connection between influence functionals and process tensors reviewed here, and using the definition of ergodicity proposed in \cite{dowling2024a}, we find that the upper bound obtained in \cite{carignano2024} is generally saturated.
This implies that influence functionals for ergodic systems are generically hard to compress as temporal matrix product states.
These findings align with the results of \cite{lu2021,ippoliti2022}, which analyze the scaling of temporal entropy in the context of the Loschmidt echoes for different classes of dynamics.

In parallel, in the open quantum systems community, there is a similarly optimistic impression that influence functionals of thermalized environments might be efficiently compressible as matrix product operators \cite{strathearn2017,cygorek2022}.
Initial results for Gaussian environments appear consistent with the observation that integrable systems yield simple environments.
An intriguing question remains whether strongly interacting, ergodic environments at equilibrium can also be easily compressible as temporal matrix product operators, which, from our perspective, has not yet been convincingly demonstrated, as discussed in \cite{park2024}.
We have also shown that extracting influence functionals requires a set of tensor network algorithms with strong connections to the density matrix renormalization group methods for open systems.
In the thermodynamic limit, these algorithms are equivalent to those used to determine the stationary state of quantum channels, which is described by the spatial transfer matrix we defined.

The temporal entropy of the separated left and right influence functionals (the leading left and right eigenvectors of such a transfer matrix)
turn out to be 
 not gauge invariant. % but can be modified via appropriate gauge transformations \cite{tang2023matrix}.
To address this, we have begun defining generalized temporal entropies that directly account for the overlap of the leading left and right eigenvectors, the left-right influence functionals \cite{Carignano_PhysRevResearch.6.033021}.
These entropies are generally complex-valued; however, in some contexts their real part exhibits only a mild growth over time \cite{carignano2024}.
This suggests the possibility of encoding the overlap of left and right influence functionals using appropriate temporal matrix product operators.
Furthermore, we have recently shown that such generalized entropies can be measured in experiments, by appropriately designed quenches on replicated systems \cite{bou2024}.
Generalized entropies thus give hope that the ergodic dynamics can still be efficiently simulated with tensor networks.

Transforming this hope into a concrete tensor network strategy is an active area of research.
Currently, one of the main obstacles is to understand how to connect the scaling of a complex-valued pseudo-entropy with the complexity of the corresponding generalized transition matrix and its compressiblily in terms of tensor networks.
Regardless of these open questions, we have reviewed here how the field of partial contractions of spatio-temporal tensor networks encoding the path integrals of many-body quantum systems out-of-equilibrium, both in open and closed scenarios, is relatively young and very promising, having 
 produced fertile connections and enhanced our understanding of out-of-equilibrium dynamics.
We thus firmly believe that this field holds great potential for further breakthroughs.

{\it Note added: } While completing this manuscript, a new preprint \cite{milekhin2025} which addresses topics related to those discussed here has appeared on the arXiv.

\section*{Funding}
We acknowledge support from the Proyecto Sinérgico CAM Y2020/TCS-6545 NanoQuCo-CM,
the CSIC Research Platform on Quantum Technologies PTI-001, and from the Grant TED2021-130552B-C22 funded by MCIN/AEI/10.13039/501100011033 and by the ``European Union NextGenerationEU/PRTR'',
and Grant PID2021-127968NB-I00 funded by MCIN/AEI/10.13039/501100011033.
ABC is supported by Grant MMT24-IFF-01 and SC by Grant C005/24-ED CV1.
The funding for these actions/grants and contracts comes from the European Union's Recovery and Resilience Facility-Next Generation, in the framework of the General Invitation of the Spanish Government's public business entity Red.es to
participate in talent attraction and retention programmes within Investment 4 of Component 19 of the Recovery, Transformation and Resilience Plan.

\section*{Acknowledgments}

We acknowledge the discussion and collaboration on topics related to those of this review with Carlos Ramos Marimon, Jacopo De Nardis and Guglielmo Lami.

\bibliography{References.bib} % Entries are in the refs.bib file

\begin{thebibliography}{62}%
\makeatletter
\providecommand \@ifxundefined [1]{%
 \@ifx{#1\undefined}
}%
\providecommand \@ifnum [1]{%
 \ifnum #1\expandafter \@firstoftwo
 \else \expandafter \@secondoftwo
 \fi
}%
\providecommand \@ifx [1]{%
 \ifx #1\expandafter \@firstoftwo
 \else \expandafter \@secondoftwo
 \fi
}%
\providecommand \natexlab [1]{#1}%
\providecommand \enquote  [1]{``#1''}%
\providecommand \bibnamefont  [1]{#1}%
\providecommand \bibfnamefont [1]{#1}%
\providecommand \citenamefont [1]{#1}%
\providecommand \href@noop [0]{\@secondoftwo}%
\providecommand \href [0]{\begingroup \@sanitize@url \@href}%
\providecommand \@href[1]{\@@startlink{#1}\@@href}%
\providecommand \@@href[1]{\endgroup#1\@@endlink}%
\providecommand \@sanitize@url [0]{\catcode `\\12\catcode `\$12\catcode
  `\&12\catcode `\#12\catcode `\^12\catcode `\_12\catcode `\%12\relax}%
\providecommand \@@startlink[1]{}%
\providecommand \@@endlink[0]{}%
\providecommand \url  [0]{\begingroup\@sanitize@url \@url }%
\providecommand \@url [1]{\endgroup\@href {#1}{\urlprefix }}%
\providecommand \urlprefix  [0]{URL }%
\providecommand \Eprint [0]{\href }%
\providecommand \doibase [0]{https://doi.org/}%
\providecommand \selectlanguage [0]{\@gobble}%
\providecommand \bibinfo  [0]{\@secondoftwo}%
\providecommand \bibfield  [0]{\@secondoftwo}%
\providecommand \translation [1]{[#1]}%
\providecommand \BibitemOpen [0]{}%
\providecommand \bibitemStop [0]{}%
\providecommand \bibitemNoStop [0]{.\EOS\space}%
\providecommand \EOS [0]{\spacefactor3000\relax}%
\providecommand \BibitemShut  [1]{\csname bibitem#1\endcsname}%
\let\auto@bib@innerbib\@empty
%</preamble>
\bibitem [{\citenamefont {Calabrese}\ and\ \citenamefont
  {Cardy}(2005)}]{calabrese_2005}%
  \BibitemOpen
  \bibfield  {author} {\bibinfo {author} {\bibfnamefont {P.}~\bibnamefont
  {Calabrese}}\ and\ \bibinfo {author} {\bibfnamefont {J.}~\bibnamefont
  {Cardy}},\ }\bibfield  {title} {\bibinfo {title} {Evolution of entanglement
  entropy in one-dimensional systems},\ }\bibfield  {journal} {\bibinfo
  {journal} {Journal of Statistical Mechanics: Theory and Experiment}\ }\href
  {https://doi.org/10.1088/1742-5468/2005/04/P04010}
  {10.1088/1742-5468/2005/04/P04010} (\bibinfo {year} {2005})\BibitemShut
  {NoStop}%
\bibitem [{\citenamefont {Läuchli}\ and\ \citenamefont
  {Kollath}(2008)}]{lauchli_spreading_2008}%
  \BibitemOpen
  \bibfield  {author} {\bibinfo {author} {\bibfnamefont {A.~M.}\ \bibnamefont
  {Läuchli}}\ and\ \bibinfo {author} {\bibfnamefont {C.}~\bibnamefont
  {Kollath}},\ }\bibfield  {title} {\bibinfo {title} {Spreading of correlations
  and entanglement after a quench in the one-dimensional {Bose}–{Hubbard}
  model},\ }\href {https://doi.org/10.1088/1742-5468/2008/05/P05018} {\bibfield
   {journal} {\bibinfo  {journal} {Journal of Statistical Mechanics: Theory and
  Experiment}\ ,\ \bibinfo {pages} {P05018}} (\bibinfo {year}
  {2008})}\BibitemShut {NoStop}%
\bibitem [{\citenamefont {Dubail}(2017)}]{dubail2017}%
  \BibitemOpen
  \bibfield  {author} {\bibinfo {author} {\bibfnamefont {J.}~\bibnamefont
  {Dubail}},\ }\bibfield  {title} {\bibinfo {title} {Entanglement scaling of
  operators: A conformal field theory approach, with a glimpse of simulability
  of long-time dynamics in 1+1d},\ }\href
  {https://doi.org/10.1088/1751-8121/aa6f38} {\bibfield  {journal} {\bibinfo
  {journal} {Journal of Physics A: Mathematical and Theoretical}\ }\textbf
  {\bibinfo {volume} {50}},\ \bibinfo {pages} {234001} (\bibinfo {year}
  {2017})},\ \Eprint {https://arxiv.org/abs/1612.08630} {arXiv:1612.08630
  [cond-mat, physics:hep-th, physics:quant-ph]} \BibitemShut {NoStop}%
\bibitem [{\citenamefont {Gottesman}(1998)}]{PhysRevA.57.127}%
  \BibitemOpen
  \bibfield  {author} {\bibinfo {author} {\bibfnamefont {D.}~\bibnamefont
  {Gottesman}},\ }\bibfield  {title} {\bibinfo {title} {Theory of
  fault-tolerant quantum computation},\ }\href
  {https://doi.org/10.1103/PhysRevA.57.127} {\bibfield  {journal} {\bibinfo
  {journal} {Phys. Rev. A}\ }\textbf {\bibinfo {volume} {57}},\ \bibinfo
  {pages} {127} (\bibinfo {year} {1998})}\BibitemShut {NoStop}%
\bibitem [{\citenamefont {Prosen}\ and\ \citenamefont {{\v Z}nidari{\v
  c}}(2007)}]{prosenEfficiencyClassicalSimulations2007}%
  \BibitemOpen
  \bibfield  {author} {\bibinfo {author} {\bibfnamefont {T.}~\bibnamefont
  {Prosen}}\ and\ \bibinfo {author} {\bibfnamefont {M.}~\bibnamefont {{\v
  Z}nidari{\v c}}},\ }\bibfield  {title} {\bibinfo {title} {Is the efficiency
  of classical simulations of quantum dynamics related to integrability?},\
  }\href {https://doi.org/10.1103/PhysRevE.75.015202} {\bibfield  {journal}
  {\bibinfo  {journal} {Physical Review E}\ }\textbf {\bibinfo {volume} {75}},\
  \bibinfo {pages} {015202} (\bibinfo {year} {2007})}\BibitemShut {NoStop}%
\bibitem [{\citenamefont {Prosen}\ and\ \citenamefont
  {Pi{\v{z}}orn}(2007)}]{prosen2007operator}%
  \BibitemOpen
  \bibfield  {author} {\bibinfo {author} {\bibfnamefont {T.}~\bibnamefont
  {Prosen}}\ and\ \bibinfo {author} {\bibfnamefont {I.}~\bibnamefont
  {Pi{\v{z}}orn}},\ }\bibfield  {title} {\bibinfo {title} {Operator space
  entanglement entropy in a transverse ising chain},\ }\href@noop {} {\bibfield
   {journal} {\bibinfo  {journal} {Physical Review A—Atomic, Molecular, and
  Optical Physics}\ }\textbf {\bibinfo {volume} {76}},\ \bibinfo {pages}
  {032316} (\bibinfo {year} {2007})}\BibitemShut {NoStop}%
\bibitem [{\citenamefont {Bertini}\ \emph
  {et~al.}(2020{\natexlab{a}})\citenamefont {Bertini}, \citenamefont {Kos},\
  and\ \citenamefont {Prosen}}]{bertiniOperatorEntanglementLocal2020}%
  \BibitemOpen
  \bibfield  {author} {\bibinfo {author} {\bibfnamefont {B.}~\bibnamefont
  {Bertini}}, \bibinfo {author} {\bibfnamefont {P.}~\bibnamefont {Kos}},\ and\
  \bibinfo {author} {\bibfnamefont {T.}~\bibnamefont {Prosen}},\ }\bibfield
  {title} {\bibinfo {title} {Operator {{Entanglement}} in {{Local Quantum
  Circuits I}}: {{Chaotic Dual-Unitary Circuits}}},\ }\href
  {https://doi.org/10.21468/SciPostPhys.8.4.067} {\bibfield  {journal}
  {\bibinfo  {journal} {SciPost Physics}\ }\textbf {\bibinfo {volume} {8}},\
  \bibinfo {pages} {067} (\bibinfo {year} {2020}{\natexlab{a}})}\BibitemShut
  {NoStop}%
\bibitem [{\citenamefont {Bertini}\ \emph
  {et~al.}(2020{\natexlab{b}})\citenamefont {Bertini}, \citenamefont {Kos},\
  and\ \citenamefont {Prosen}}]{bertiniOperatorEntanglementLocal2020a}%
  \BibitemOpen
  \bibfield  {author} {\bibinfo {author} {\bibfnamefont {B.}~\bibnamefont
  {Bertini}}, \bibinfo {author} {\bibfnamefont {P.}~\bibnamefont {Kos}},\ and\
  \bibinfo {author} {\bibfnamefont {T.}~\bibnamefont {Prosen}},\ }\bibfield
  {title} {\bibinfo {title} {Operator {{Entanglement}} in {{Local Quantum
  Circuits II}}: {{Solitons}} in {{Chains}} of {{Qubits}}},\ }\href
  {https://doi.org/10.21468/SciPostPhys.8.4.068} {\bibfield  {journal}
  {\bibinfo  {journal} {SciPost Physics}\ }\textbf {\bibinfo {volume} {8}},\
  \bibinfo {pages} {068} (\bibinfo {year} {2020}{\natexlab{b}})}\BibitemShut
  {NoStop}%
\bibitem [{\citenamefont {Giudice}\ \emph {et~al.}(2022)\citenamefont
  {Giudice}, \citenamefont {Giudici}, \citenamefont {Sonner}, \citenamefont
  {Thoenniss}, \citenamefont {Lerose}, \citenamefont {Abanin},\ and\
  \citenamefont {Piroli}}]{Giudice_PhysRevLett.128.220401}%
  \BibitemOpen
  \bibfield  {author} {\bibinfo {author} {\bibfnamefont {G.}~\bibnamefont
  {Giudice}}, \bibinfo {author} {\bibfnamefont {G.}~\bibnamefont {Giudici}},
  \bibinfo {author} {\bibfnamefont {M.}~\bibnamefont {Sonner}}, \bibinfo
  {author} {\bibfnamefont {J.}~\bibnamefont {Thoenniss}}, \bibinfo {author}
  {\bibfnamefont {A.}~\bibnamefont {Lerose}}, \bibinfo {author} {\bibfnamefont
  {D.~A.}\ \bibnamefont {Abanin}},\ and\ \bibinfo {author} {\bibfnamefont
  {L.}~\bibnamefont {Piroli}},\ }\bibfield  {title} {\bibinfo {title} {Temporal
  entanglement, quasiparticles, and the role of interactions},\ }\href
  {https://doi.org/10.1103/PhysRevLett.128.220401} {\bibfield  {journal}
  {\bibinfo  {journal} {Phys. Rev. Lett.}\ }\textbf {\bibinfo {volume} {128}},\
  \bibinfo {pages} {220401} (\bibinfo {year} {2022})}\BibitemShut {NoStop}%
\bibitem [{\citenamefont {Thoenniss}\ \emph {et~al.}(2023)\citenamefont
  {Thoenniss}, \citenamefont {Lerose},\ and\ \citenamefont
  {Abanin}}]{Thoenniss_PhysRevB.107.195101}%
  \BibitemOpen
  \bibfield  {author} {\bibinfo {author} {\bibfnamefont {J.}~\bibnamefont
  {Thoenniss}}, \bibinfo {author} {\bibfnamefont {A.}~\bibnamefont {Lerose}},\
  and\ \bibinfo {author} {\bibfnamefont {D.~A.}\ \bibnamefont {Abanin}},\
  }\bibfield  {title} {\bibinfo {title} {Nonequilibrium quantum impurity
  problems via matrix-product states in the temporal domain},\ }\href
  {https://doi.org/10.1103/PhysRevB.107.195101} {\bibfield  {journal} {\bibinfo
   {journal} {Phys. Rev. B}\ }\textbf {\bibinfo {volume} {107}},\ \bibinfo
  {pages} {195101} (\bibinfo {year} {2023})}\BibitemShut {NoStop}%
\bibitem [{\citenamefont {Ba{\~n}uls}\ \emph {et~al.}(2009)\citenamefont
  {Ba{\~n}uls}, \citenamefont {Hastings}, \citenamefont {Verstraete},\ and\
  \citenamefont {Cirac}}]{banuls2009}%
  \BibitemOpen
  \bibfield  {author} {\bibinfo {author} {\bibfnamefont {M.~C.}\ \bibnamefont
  {Ba{\~n}uls}}, \bibinfo {author} {\bibfnamefont {M.~B.}\ \bibnamefont
  {Hastings}}, \bibinfo {author} {\bibfnamefont {F.}~\bibnamefont
  {Verstraete}},\ and\ \bibinfo {author} {\bibfnamefont {J.~I.}\ \bibnamefont
  {Cirac}},\ }\bibfield  {title} {\bibinfo {title} {Matrix {{Product States}}
  for {{Dynamical Simulation}} of {{Infinite Chains}}},\ }\href
  {https://doi.org/10.1103/PhysRevLett.102.240603} {\bibfield  {journal}
  {\bibinfo  {journal} {Physical Review Letters}\ }\textbf {\bibinfo {volume}
  {102}},\ \bibinfo {pages} {240603} (\bibinfo {year} {2009})}\BibitemShut
  {NoStop}%
\bibitem [{\citenamefont {{M{\"u}ller-Hermes}}\ \emph
  {et~al.}(2012)\citenamefont {{M{\"u}ller-Hermes}}, \citenamefont {Cirac},\
  and\ \citenamefont {Ba{\~n}uls}}]{muller-hermes2012}%
  \BibitemOpen
  \bibfield  {author} {\bibinfo {author} {\bibfnamefont {A.}~\bibnamefont
  {{M{\"u}ller-Hermes}}}, \bibinfo {author} {\bibfnamefont {J.~I.}\
  \bibnamefont {Cirac}},\ and\ \bibinfo {author} {\bibfnamefont {M.~C.}\
  \bibnamefont {Ba{\~n}uls}},\ }\bibfield  {title} {\bibinfo {title} {Tensor
  network techniques for the computation of dynamical observables in {{1D}}
  quantum spin systems},\ }\href
  {https://doi.org/10.1088/1367-2630/14/7/075003} {\bibfield  {journal}
  {\bibinfo  {journal} {New Journal of Physics}\ }\textbf {\bibinfo {volume}
  {14}},\ \bibinfo {pages} {075003} (\bibinfo {year} {2012})},\ \Eprint
  {https://arxiv.org/abs/1204.5080} {arXiv:1204.5080 [cond-mat,
  physics:quant-ph]} \BibitemShut {NoStop}%
\bibitem [{\citenamefont {Surace}\ \emph {et~al.}(2019)\citenamefont {Surace},
  \citenamefont {Piani},\ and\ \citenamefont {Tagliacozzo}}]{surace2019a}%
  \BibitemOpen
  \bibfield  {author} {\bibinfo {author} {\bibfnamefont {J.}~\bibnamefont
  {Surace}}, \bibinfo {author} {\bibfnamefont {M.}~\bibnamefont {Piani}},\ and\
  \bibinfo {author} {\bibfnamefont {L.}~\bibnamefont {Tagliacozzo}},\
  }\bibfield  {title} {\bibinfo {title} {Simulating the out-of-equilibrium
  dynamics of local observables by trading entanglement for mixture},\ }\href
  {https://doi.org/10.1103/PhysRevB.99.235115} {\bibfield  {journal} {\bibinfo
  {journal} {Physical Review B}\ }\textbf {\bibinfo {volume} {99}},\ \bibinfo
  {pages} {235115} (\bibinfo {year} {2019})}\BibitemShut {NoStop}%
\bibitem [{\citenamefont {White}\ \emph {et~al.}(2018)\citenamefont {White},
  \citenamefont {Zaletel}, \citenamefont {Mong},\ and\ \citenamefont
  {Refael}}]{white2018}%
  \BibitemOpen
  \bibfield  {author} {\bibinfo {author} {\bibfnamefont {C.~D.}\ \bibnamefont
  {White}}, \bibinfo {author} {\bibfnamefont {M.}~\bibnamefont {Zaletel}},
  \bibinfo {author} {\bibfnamefont {R.~S.~K.}\ \bibnamefont {Mong}},\ and\
  \bibinfo {author} {\bibfnamefont {G.}~\bibnamefont {Refael}},\ }\bibfield
  {title} {\bibinfo {title} {Quantum dynamics of thermalizing systems},\ }\href
  {https://doi.org/10.1103/PhysRevB.97.035127} {\bibfield  {journal} {\bibinfo
  {journal} {Physical Review B}\ }\textbf {\bibinfo {volume} {97}},\ \bibinfo
  {pages} {035127} (\bibinfo {year} {2018})},\ \Eprint
  {https://arxiv.org/abs/1707.01506} {arXiv:1707.01506} \BibitemShut {NoStop}%
\bibitem [{\citenamefont {Fr\'{\i}as-P\'erez}\ and\ \citenamefont
  {Ba\~nuls}(2022)}]{frias2022}%
  \BibitemOpen
  \bibfield  {author} {\bibinfo {author} {\bibfnamefont {M.}~\bibnamefont
  {Fr\'{\i}as-P\'erez}}\ and\ \bibinfo {author} {\bibfnamefont {M.~C.}\
  \bibnamefont {Ba\~nuls}},\ }\bibfield  {title} {\bibinfo {title} {Light cone
  tensor network and time evolution},\ }\href
  {https://doi.org/10.1103/PhysRevB.106.115117} {\bibfield  {journal} {\bibinfo
   {journal} {Phys. Rev. B}\ }\textbf {\bibinfo {volume} {106}},\ \bibinfo
  {pages} {115117} (\bibinfo {year} {2022})}\BibitemShut {NoStop}%
\bibitem [{\citenamefont {Paeckel}\ \emph {et~al.}(2019)\citenamefont
  {Paeckel}, \citenamefont {K{\"o}hler}, \citenamefont {Swoboda}, \citenamefont
  {Manmana}, \citenamefont {Schollw{\"o}ck},\ and\ \citenamefont
  {Hubig}}]{paeckel2019time}%
  \BibitemOpen
  \bibfield  {author} {\bibinfo {author} {\bibfnamefont {S.}~\bibnamefont
  {Paeckel}}, \bibinfo {author} {\bibfnamefont {T.}~\bibnamefont {K{\"o}hler}},
  \bibinfo {author} {\bibfnamefont {A.}~\bibnamefont {Swoboda}}, \bibinfo
  {author} {\bibfnamefont {S.~R.}\ \bibnamefont {Manmana}}, \bibinfo {author}
  {\bibfnamefont {U.}~\bibnamefont {Schollw{\"o}ck}},\ and\ \bibinfo {author}
  {\bibfnamefont {C.}~\bibnamefont {Hubig}},\ }\bibfield  {title} {\bibinfo
  {title} {Time-evolution methods for matrix-product states},\ }\href@noop {}
  {\bibfield  {journal} {\bibinfo  {journal} {Annals of Physics}\ }\textbf
  {\bibinfo {volume} {411}},\ \bibinfo {pages} {167998} (\bibinfo {year}
  {2019})}\BibitemShut {NoStop}%
\bibitem [{\citenamefont {Rivas}\ \emph {et~al.}(2010)\citenamefont {Rivas},
  \citenamefont {Huelga},\ and\ \citenamefont {Plenio}}]{rivas2010}%
  \BibitemOpen
  \bibfield  {author} {\bibinfo {author} {\bibfnamefont {{\'A}.}~\bibnamefont
  {Rivas}}, \bibinfo {author} {\bibfnamefont {S.~F.}\ \bibnamefont {Huelga}},\
  and\ \bibinfo {author} {\bibfnamefont {M.~B.}\ \bibnamefont {Plenio}},\
  }\bibfield  {title} {\bibinfo {title} {Entanglement and {{Non-Markovianity}}
  of {{Quantum Evolutions}}},\ }\href
  {https://doi.org/10.1103/PhysRevLett.105.050403} {\bibfield  {journal}
  {\bibinfo  {journal} {Physical Review Letters}\ }\textbf {\bibinfo {volume}
  {105}},\ \bibinfo {pages} {050403} (\bibinfo {year} {2010})}\BibitemShut
  {NoStop}%
\bibitem [{\citenamefont {Dowling}\ \emph {et~al.}(2024)\citenamefont
  {Dowling}, \citenamefont {Modi}, \citenamefont {Mu{\~n}oz}, \citenamefont
  {Singh},\ and\ \citenamefont {White}}]{dowling2024}%
  \BibitemOpen
  \bibfield  {author} {\bibinfo {author} {\bibfnamefont {N.}~\bibnamefont
  {Dowling}}, \bibinfo {author} {\bibfnamefont {K.}~\bibnamefont {Modi}},
  \bibinfo {author} {\bibfnamefont {R.~N.}\ \bibnamefont {Mu{\~n}oz}}, \bibinfo
  {author} {\bibfnamefont {S.}~\bibnamefont {Singh}},\ and\ \bibinfo {author}
  {\bibfnamefont {G.~A.~L.}\ \bibnamefont {White}},\ }\href
  {https://doi.org/10.48550/arXiv.2312.04624} {\bibinfo {title} {Capturing
  long-range memory structures with tree-geometry process tensors}} (\bibinfo
  {year} {2024}),\ \Eprint {https://arxiv.org/abs/2312.04624}
  {arXiv:2312.04624} \BibitemShut {NoStop}%
\bibitem [{\citenamefont {Feynman}\ and\ \citenamefont
  {Vernon}(1963)}]{feynman1963}%
  \BibitemOpen
  \bibfield  {author} {\bibinfo {author} {\bibfnamefont {R.}~\bibnamefont
  {Feynman}}\ and\ \bibinfo {author} {\bibfnamefont {F.}~\bibnamefont
  {Vernon}},\ }\bibfield  {title} {\bibinfo {title} {The theory of a general
  quantum system interacting with a linear dissipative system},\ }\href
  {https://doi.org/10.1016/0003-4916(63)90068-X} {\bibfield  {journal}
  {\bibinfo  {journal} {Annals of Physics}\ }\textbf {\bibinfo {volume} {24}},\
  \bibinfo {pages} {118} (\bibinfo {year} {1963})}\BibitemShut {NoStop}%
\bibitem [{\citenamefont {Chiribella}\ \emph {et~al.}(2008)\citenamefont
  {Chiribella}, \citenamefont {D'Ariano},\ and\ \citenamefont
  {Perinotti}}]{chiribella2008}%
  \BibitemOpen
  \bibfield  {author} {\bibinfo {author} {\bibfnamefont {G.}~\bibnamefont
  {Chiribella}}, \bibinfo {author} {\bibfnamefont {G.~M.}\ \bibnamefont
  {D'Ariano}},\ and\ \bibinfo {author} {\bibfnamefont {P.}~\bibnamefont
  {Perinotti}},\ }\bibfield  {title} {\bibinfo {title} {Quantum {{Circuit
  Architecture}}},\ }\href {https://doi.org/10.1103/PhysRevLett.101.060401}
  {\bibfield  {journal} {\bibinfo  {journal} {Physical Review Letters}\
  }\textbf {\bibinfo {volume} {101}},\ \bibinfo {pages} {060401} (\bibinfo
  {year} {2008})}\BibitemShut {NoStop}%
\bibitem [{\citenamefont {Hastings}\ and\ \citenamefont
  {Mahajan}(2015)}]{hastings2015}%
  \BibitemOpen
  \bibfield  {author} {\bibinfo {author} {\bibfnamefont {M.~B.}\ \bibnamefont
  {Hastings}}\ and\ \bibinfo {author} {\bibfnamefont {R.}~\bibnamefont
  {Mahajan}},\ }\bibfield  {title} {\bibinfo {title} {Connecting
  {{Entanglement}} in {{Time}} and {{Space}}: {{Improving}} the {{Folding
  Algorithm}}},\ }\href {https://doi.org/10.1103/PhysRevA.91.032306} {\bibfield
   {journal} {\bibinfo  {journal} {Physical Review A}\ }\textbf {\bibinfo
  {volume} {91}},\ \bibinfo {pages} {032306} (\bibinfo {year} {2015})},\
  \Eprint {https://arxiv.org/abs/1411.7950} {arXiv:1411.7950 [cond-mat,
  physics:hep-th, physics:quant-ph]} \BibitemShut {NoStop}%
\bibitem [{\citenamefont {Carignano}\ \emph {et~al.}(2024)\citenamefont
  {Carignano}, \citenamefont {Marim\'on},\ and\ \citenamefont
  {Tagliacozzo}}]{Carignano_PhysRevResearch.6.033021}%
  \BibitemOpen
  \bibfield  {author} {\bibinfo {author} {\bibfnamefont {S.}~\bibnamefont
  {Carignano}}, \bibinfo {author} {\bibfnamefont {C.~R.}\ \bibnamefont
  {Marim\'on}},\ and\ \bibinfo {author} {\bibfnamefont {L.}~\bibnamefont
  {Tagliacozzo}},\ }\bibfield  {title} {\bibinfo {title} {Temporal entropy and
  the complexity of computing the expectation value of local operators after a
  quench},\ }\href {https://doi.org/10.1103/PhysRevResearch.6.033021}
  {\bibfield  {journal} {\bibinfo  {journal} {Phys. Rev. Res.}\ }\textbf
  {\bibinfo {volume} {6}},\ \bibinfo {pages} {033021} (\bibinfo {year}
  {2024})}\BibitemShut {NoStop}%
\bibitem [{\citenamefont {Vidal}(2004)}]{vidal2004efficient}%
  \BibitemOpen
  \bibfield  {author} {\bibinfo {author} {\bibfnamefont {G.}~\bibnamefont
  {Vidal}},\ }\bibfield  {title} {\bibinfo {title} {Efficient simulation of
  one-dimensional quantum many-body systems},\ }\href@noop {} {\bibfield
  {journal} {\bibinfo  {journal} {Physical review letters}\ }\textbf {\bibinfo
  {volume} {93}},\ \bibinfo {pages} {040502} (\bibinfo {year}
  {2004})}\BibitemShut {NoStop}%
\bibitem [{\citenamefont {Tirrito}\ \emph {et~al.}(2018)\citenamefont
  {Tirrito}, \citenamefont {Robinson}, \citenamefont {Lewenstein},
  \citenamefont {Ran},\ and\ \citenamefont
  {Tagliacozzo}}]{tirrito2018characterizing}%
  \BibitemOpen
  \bibfield  {author} {\bibinfo {author} {\bibfnamefont {E.}~\bibnamefont
  {Tirrito}}, \bibinfo {author} {\bibfnamefont {N.~J.}\ \bibnamefont
  {Robinson}}, \bibinfo {author} {\bibfnamefont {M.}~\bibnamefont
  {Lewenstein}}, \bibinfo {author} {\bibfnamefont {S.-J.}\ \bibnamefont
  {Ran}},\ and\ \bibinfo {author} {\bibfnamefont {L.}~\bibnamefont
  {Tagliacozzo}},\ }\bibfield  {title} {\bibinfo {title} {Characterizing the
  quantum field theory vacuum using temporal matrix product states},\
  }\href@noop {} {\bibfield  {journal} {\bibinfo  {journal} {arXiv preprint
  arXiv:1810.08050}\ } (\bibinfo {year} {2018})}\BibitemShut {NoStop}%
\bibitem [{\citenamefont {Schuch}\ \emph {et~al.}(2007)\citenamefont {Schuch},
  \citenamefont {Wolf}, \citenamefont {Verstraete},\ and\ \citenamefont
  {Cirac}}]{schuch2007computational}%
  \BibitemOpen
  \bibfield  {author} {\bibinfo {author} {\bibfnamefont {N.}~\bibnamefont
  {Schuch}}, \bibinfo {author} {\bibfnamefont {M.~M.}\ \bibnamefont {Wolf}},
  \bibinfo {author} {\bibfnamefont {F.}~\bibnamefont {Verstraete}},\ and\
  \bibinfo {author} {\bibfnamefont {J.~I.}\ \bibnamefont {Cirac}},\ }\bibfield
  {title} {\bibinfo {title} {Computational complexity of projected entangled
  pair states},\ }\href@noop {} {\bibfield  {journal} {\bibinfo  {journal}
  {Physical review letters}\ }\textbf {\bibinfo {volume} {98}},\ \bibinfo
  {pages} {140506} (\bibinfo {year} {2007})}\BibitemShut {NoStop}%
\bibitem [{\citenamefont {Verstraete}\ and\ \citenamefont
  {Cirac}(2006)}]{verstraete2006}%
  \BibitemOpen
  \bibfield  {author} {\bibinfo {author} {\bibfnamefont {F.}~\bibnamefont
  {Verstraete}}\ and\ \bibinfo {author} {\bibfnamefont {J.~I.}\ \bibnamefont
  {Cirac}},\ }\bibfield  {title} {\bibinfo {title} {Matrix product states
  represent ground states faithfully},\ }\href
  {https://doi.org/10.1103/PhysRevB.73.094423} {\bibfield  {journal} {\bibinfo
  {journal} {Physical Review B}\ }\textbf {\bibinfo {volume} {73}},\ \bibinfo
  {pages} {094423} (\bibinfo {year} {2006})}\BibitemShut {NoStop}%
\bibitem [{\citenamefont {Pi{\v z}orn}\ and\ \citenamefont
  {Prosen}(2009)}]{pizornOperatorSpaceEntanglement2009}%
  \BibitemOpen
  \bibfield  {author} {\bibinfo {author} {\bibfnamefont {I.}~\bibnamefont
  {Pi{\v z}orn}}\ and\ \bibinfo {author} {\bibfnamefont {T.}~\bibnamefont
  {Prosen}},\ }\bibfield  {title} {\bibinfo {title} {Operator space
  entanglement entropy in \${{XY}}\$ spin chains},\ }\href
  {https://doi.org/10.1103/PhysRevB.79.184416} {\bibfield  {journal} {\bibinfo
  {journal} {Physical Review B}\ }\textbf {\bibinfo {volume} {79}},\ \bibinfo
  {pages} {184416} (\bibinfo {year} {2009})}\BibitemShut {NoStop}%
\bibitem [{\citenamefont {Lerose}\ \emph {et~al.}(2023)\citenamefont {Lerose},
  \citenamefont {Sonner},\ and\ \citenamefont {Abanin}}]{lerose2023overcoming}%
  \BibitemOpen
  \bibfield  {author} {\bibinfo {author} {\bibfnamefont {A.}~\bibnamefont
  {Lerose}}, \bibinfo {author} {\bibfnamefont {M.}~\bibnamefont {Sonner}},\
  and\ \bibinfo {author} {\bibfnamefont {D.~A.}\ \bibnamefont {Abanin}},\
  }\bibfield  {title} {\bibinfo {title} {Overcoming the entanglement barrier in
  quantum many-body dynamics via space-time duality},\ }\href@noop {}
  {\bibfield  {journal} {\bibinfo  {journal} {Physical Review B}\ }\textbf
  {\bibinfo {volume} {107}},\ \bibinfo {pages} {L060305} (\bibinfo {year}
  {2023})}\BibitemShut {NoStop}%
\bibitem [{\citenamefont {Ippoliti}\ and\ \citenamefont
  {Khemani}(2021)}]{ippoliti2021}%
  \BibitemOpen
  \bibfield  {author} {\bibinfo {author} {\bibfnamefont {M.}~\bibnamefont
  {Ippoliti}}\ and\ \bibinfo {author} {\bibfnamefont {V.}~\bibnamefont
  {Khemani}},\ }\bibfield  {title} {\bibinfo {title} {Postselection-{{Free
  Entanglement Dynamics}} via {{Spacetime Duality}}},\ }\href
  {https://doi.org/10.1103/PhysRevLett.126.060501} {\bibfield  {journal}
  {\bibinfo  {journal} {Physical Review Letters}\ }\textbf {\bibinfo {volume}
  {126}},\ \bibinfo {pages} {060501} (\bibinfo {year} {2021})}\BibitemShut
  {NoStop}%
\bibitem [{\citenamefont {Petrat}\ and\ \citenamefont
  {Tumulka}(2014)}]{petrat2014multi}%
  \BibitemOpen
  \bibfield  {author} {\bibinfo {author} {\bibfnamefont {S.}~\bibnamefont
  {Petrat}}\ and\ \bibinfo {author} {\bibfnamefont {R.}~\bibnamefont
  {Tumulka}},\ }\bibfield  {title} {\bibinfo {title} {Multi-time wave functions
  for quantum field theory},\ }\href@noop {} {\bibfield  {journal} {\bibinfo
  {journal} {Annals of Physics}\ }\textbf {\bibinfo {volume} {345}},\ \bibinfo
  {pages} {17} (\bibinfo {year} {2014})}\BibitemShut {NoStop}%
\bibitem [{\citenamefont {Lerose}\ \emph
  {et~al.}(2021{\natexlab{a}})\citenamefont {Lerose}, \citenamefont {Sonner},\
  and\ \citenamefont {Abanin}}]{lerose2021influence}%
  \BibitemOpen
  \bibfield  {author} {\bibinfo {author} {\bibfnamefont {A.}~\bibnamefont
  {Lerose}}, \bibinfo {author} {\bibfnamefont {M.}~\bibnamefont {Sonner}},\
  and\ \bibinfo {author} {\bibfnamefont {D.~A.}\ \bibnamefont {Abanin}},\
  }\bibfield  {title} {\bibinfo {title} {Influence matrix approach to many-body
  floquet dynamics},\ }\href@noop {} {\bibfield  {journal} {\bibinfo  {journal}
  {Physical Review X}\ }\textbf {\bibinfo {volume} {11}},\ \bibinfo {pages}
  {021040} (\bibinfo {year} {2021}{\natexlab{a}})}\BibitemShut {NoStop}%
\bibitem [{\citenamefont {Chiribella}\ \emph {et~al.}(2009)\citenamefont
  {Chiribella}, \citenamefont {D'Ariano},\ and\ \citenamefont
  {Perinotti}}]{chiribella2009}%
  \BibitemOpen
  \bibfield  {author} {\bibinfo {author} {\bibfnamefont {G.}~\bibnamefont
  {Chiribella}}, \bibinfo {author} {\bibfnamefont {G.~M.}\ \bibnamefont
  {D'Ariano}},\ and\ \bibinfo {author} {\bibfnamefont {P.}~\bibnamefont
  {Perinotti}},\ }\bibfield  {title} {\bibinfo {title} {Theoretical framework
  for quantum networks},\ }\href {https://doi.org/10.1103/PhysRevA.80.022339}
  {\bibfield  {journal} {\bibinfo  {journal} {Physical Review A}\ }\textbf
  {\bibinfo {volume} {80}},\ \bibinfo {pages} {022339} (\bibinfo {year}
  {2009})}\BibitemShut {NoStop}%
\bibitem [{\citenamefont {Aharonov}\ \emph {et~al.}(2007)\citenamefont
  {Aharonov}, \citenamefont {Popescu}, \citenamefont {Tollaksen},\ and\
  \citenamefont {Vaidman}}]{aharonov2007}%
  \BibitemOpen
  \bibfield  {author} {\bibinfo {author} {\bibfnamefont {Y.}~\bibnamefont
  {Aharonov}}, \bibinfo {author} {\bibfnamefont {S.}~\bibnamefont {Popescu}},
  \bibinfo {author} {\bibfnamefont {J.}~\bibnamefont {Tollaksen}},\ and\
  \bibinfo {author} {\bibfnamefont {L.}~\bibnamefont {Vaidman}},\ }\href
  {https://doi.org/10.1103/PhysRevA.79.052110} {\bibinfo {title} {Multiple-time
  states and multiple-time measurements in quantum mechanics}} (\bibinfo {year}
  {2007}),\ \Eprint {https://arxiv.org/abs/0712.0320} {arXiv:0712.0320
  [quant-ph]} \BibitemShut {NoStop}%
\bibitem [{\citenamefont {Leifer}(2006)}]{leifer2006}%
  \BibitemOpen
  \bibfield  {author} {\bibinfo {author} {\bibfnamefont {M.~S.}\ \bibnamefont
  {Leifer}},\ }\bibfield  {title} {\bibinfo {title} {Quantum dynamics as an
  analog of conditional probability},\ }\bibfield  {journal} {\bibinfo
  {journal} {Physical Review A}\ }\textbf {\bibinfo {volume} {74}},\ \href
  {https://doi.org/10.1103/PhysRevA.74.042310} {10.1103/PhysRevA.74.042310}
  (\bibinfo {year} {2006})\BibitemShut {NoStop}%
\bibitem [{\citenamefont {Leifer}\ and\ \citenamefont
  {Spekkens}(2013)}]{leifer2013}%
  \BibitemOpen
  \bibfield  {author} {\bibinfo {author} {\bibfnamefont {M.~S.}\ \bibnamefont
  {Leifer}}\ and\ \bibinfo {author} {\bibfnamefont {R.~W.}\ \bibnamefont
  {Spekkens}},\ }\bibfield  {title} {\bibinfo {title} {Towards a formulation of
  quantum theory as a causally neutral theory of {{Bayesian}} inference},\
  }\href {https://doi.org/10.1103/PhysRevA.88.052130} {\bibfield  {journal}
  {\bibinfo  {journal} {Physical Review A}\ }\textbf {\bibinfo {volume} {88}},\
  \bibinfo {pages} {052130} (\bibinfo {year} {2013})}\BibitemShut {NoStop}%
\bibitem [{\citenamefont {Pollock}\ \emph {et~al.}(2018)\citenamefont
  {Pollock}, \citenamefont {Rodr\'{\i}guez-Rosario}, \citenamefont
  {Frauenheim}, \citenamefont {Paternostro},\ and\ \citenamefont
  {Modi}}]{pollock2018}%
  \BibitemOpen
  \bibfield  {author} {\bibinfo {author} {\bibfnamefont {F.~A.}\ \bibnamefont
  {Pollock}}, \bibinfo {author} {\bibfnamefont {C.}~\bibnamefont
  {Rodr\'{\i}guez-Rosario}}, \bibinfo {author} {\bibfnamefont {T.}~\bibnamefont
  {Frauenheim}}, \bibinfo {author} {\bibfnamefont {M.}~\bibnamefont
  {Paternostro}},\ and\ \bibinfo {author} {\bibfnamefont {K.}~\bibnamefont
  {Modi}},\ }\bibfield  {title} {\bibinfo {title} {Operational markov condition
  for quantum processes},\ }\href
  {https://doi.org/10.1103/PhysRevLett.120.040405} {\bibfield  {journal}
  {\bibinfo  {journal} {Phys. Rev. Lett.}\ }\textbf {\bibinfo {volume} {120}},\
  \bibinfo {pages} {040405} (\bibinfo {year} {2018})}\BibitemShut {NoStop}%
\bibitem [{\citenamefont {Pollock}(2018)}]{pollock2018a}%
  \BibitemOpen
  \bibfield  {author} {\bibinfo {author} {\bibfnamefont {F.~A.}\ \bibnamefont
  {Pollock}},\ }\bibfield  {title} {\bibinfo {title} {Non-{{Markovian}} quantum
  processes: {{Complete}} framework and efficient characterization},\
  }\bibfield  {journal} {\bibinfo  {journal} {Physical Review A}\ }\textbf
  {\bibinfo {volume} {97}},\ \href {https://doi.org/10.1103/PhysRevA.97.012127}
  {10.1103/PhysRevA.97.012127} (\bibinfo {year} {2018})\BibitemShut {NoStop}%
\bibitem [{\citenamefont {Nakata}\ \emph {et~al.}(2021)\citenamefont {Nakata},
  \citenamefont {Takayanagi}, \citenamefont {Taki}, \citenamefont {Tamaoka},\
  and\ \citenamefont {Wei}}]{nakata2021}%
  \BibitemOpen
  \bibfield  {author} {\bibinfo {author} {\bibfnamefont {Y.}~\bibnamefont
  {Nakata}}, \bibinfo {author} {\bibfnamefont {T.}~\bibnamefont {Takayanagi}},
  \bibinfo {author} {\bibfnamefont {Y.}~\bibnamefont {Taki}}, \bibinfo {author}
  {\bibfnamefont {K.}~\bibnamefont {Tamaoka}},\ and\ \bibinfo {author}
  {\bibfnamefont {Z.}~\bibnamefont {Wei}},\ }\bibfield  {title} {\bibinfo
  {title} {Holographic {{Pseudo Entropy}}},\ }\href
  {https://doi.org/10.1103/PhysRevD.103.026005} {\bibfield  {journal} {\bibinfo
   {journal} {Physical Review D}\ }\textbf {\bibinfo {volume} {103}},\ \bibinfo
  {pages} {026005} (\bibinfo {year} {2021})},\ \Eprint
  {https://arxiv.org/abs/2005.13801} {arXiv:2005.13801 [cond-mat,
  physics:hep-th, physics:quant-ph]} \BibitemShut {NoStop}%
\bibitem [{\citenamefont {Doi}\ \emph {et~al.}(2023)\citenamefont {Doi},
  \citenamefont {Harper}, \citenamefont {Mollabashi}, \citenamefont
  {Takayanagi},\ and\ \citenamefont {Taki}}]{doi2023a}%
  \BibitemOpen
  \bibfield  {author} {\bibinfo {author} {\bibfnamefont {K.}~\bibnamefont
  {Doi}}, \bibinfo {author} {\bibfnamefont {J.}~\bibnamefont {Harper}},
  \bibinfo {author} {\bibfnamefont {A.}~\bibnamefont {Mollabashi}}, \bibinfo
  {author} {\bibfnamefont {T.}~\bibnamefont {Takayanagi}},\ and\ \bibinfo
  {author} {\bibfnamefont {Y.}~\bibnamefont {Taki}},\ }\bibfield  {title}
  {\bibinfo {title} {Pseudo {{Entropy}} in {{dS}}/{{CFT}} and {{Time-like
  Entanglement Entropy}}},\ }\href
  {https://doi.org/10.1103/PhysRevLett.130.031601} {\bibfield  {journal}
  {\bibinfo  {journal} {Physical Review Letters}\ }\textbf {\bibinfo {volume}
  {130}},\ \bibinfo {pages} {031601} (\bibinfo {year} {2023})},\ \Eprint
  {https://arxiv.org/abs/2210.09457} {arXiv:2210.09457 [cond-mat,
  physics:hep-th, physics:quant-ph]} \BibitemShut {NoStop}%
\bibitem [{\citenamefont {Carignano}\ and\ \citenamefont
  {Tagliacozzo}(2024)}]{carignano2024}%
  \BibitemOpen
  \bibfield  {author} {\bibinfo {author} {\bibfnamefont {S.}~\bibnamefont
  {Carignano}}\ and\ \bibinfo {author} {\bibfnamefont {L.}~\bibnamefont
  {Tagliacozzo}},\ }\href {https://doi.org/10.48550/arXiv.2405.14706} {\bibinfo
  {title} {Loschmidt echo, emerging dual unitarity and scaling of generalized
  temporal entropies after quenches to the critical point}} (\bibinfo {year}
  {2024}),\ \Eprint {https://arxiv.org/abs/2405.14706} {arXiv:2405.14706
  [cond-mat]} \BibitemShut {NoStop}%
\bibitem [{\citenamefont {Tang}\ \emph {et~al.}(2025)\citenamefont {Tang},
  \citenamefont {Verstraete},\ and\ \citenamefont {Haegeman}}]{tang2023matrix}%
  \BibitemOpen
  \bibfield  {author} {\bibinfo {author} {\bibfnamefont {W.}~\bibnamefont
  {Tang}}, \bibinfo {author} {\bibfnamefont {F.}~\bibnamefont {Verstraete}},\
  and\ \bibinfo {author} {\bibfnamefont {J.}~\bibnamefont {Haegeman}},\
  }\bibfield  {title} {\bibinfo {title} {Matrix product state fixed points of
  non-hermitian transfer matrices},\ }\href
  {https://doi.org/10.1103/PhysRevB.111.035107} {\bibfield  {journal} {\bibinfo
   {journal} {Phys. Rev. B}\ }\textbf {\bibinfo {volume} {111}},\ \bibinfo
  {pages} {035107} (\bibinfo {year} {2025})}\BibitemShut {NoStop}%
\bibitem [{\citenamefont {Wang}\ and\ \citenamefont
  {Xiang}(1997)}]{wang1997transfer}%
  \BibitemOpen
  \bibfield  {author} {\bibinfo {author} {\bibfnamefont {X.}~\bibnamefont
  {Wang}}\ and\ \bibinfo {author} {\bibfnamefont {T.}~\bibnamefont {Xiang}},\
  }\bibfield  {title} {\bibinfo {title} {Transfer-matrix density-matrix
  renormalization-group theory for thermodynamics of one-dimensional quantum
  systems},\ }\href@noop {} {\bibfield  {journal} {\bibinfo  {journal}
  {Physical Review B}\ }\textbf {\bibinfo {volume} {56}},\ \bibinfo {pages}
  {5061} (\bibinfo {year} {1997})}\BibitemShut {NoStop}%
\bibitem [{\citenamefont {Xiang}(1998)}]{xiang1998thermodynamics}%
  \BibitemOpen
  \bibfield  {author} {\bibinfo {author} {\bibfnamefont {T.}~\bibnamefont
  {Xiang}},\ }\bibfield  {title} {\bibinfo {title} {Thermodynamics of quantum
  heisenberg spin chains},\ }\href@noop {} {\bibfield  {journal} {\bibinfo
  {journal} {Physical Review B}\ }\textbf {\bibinfo {volume} {58}},\ \bibinfo
  {pages} {9142} (\bibinfo {year} {1998})}\BibitemShut {NoStop}%
\bibitem [{\citenamefont {Zhong}\ \emph {et~al.}(2024)\citenamefont {Zhong},
  \citenamefont {Pan}, \citenamefont {Lin}, \citenamefont {Wang},\ and\
  \citenamefont {Hu}}]{zhong2024density}%
  \BibitemOpen
  \bibfield  {author} {\bibinfo {author} {\bibfnamefont {P.}~\bibnamefont
  {Zhong}}, \bibinfo {author} {\bibfnamefont {W.}~\bibnamefont {Pan}}, \bibinfo
  {author} {\bibfnamefont {H.}~\bibnamefont {Lin}}, \bibinfo {author}
  {\bibfnamefont {X.}~\bibnamefont {Wang}},\ and\ \bibinfo {author}
  {\bibfnamefont {S.}~\bibnamefont {Hu}},\ }\bibfield  {title} {\bibinfo
  {title} {Density-matrix renormalization group algorithm for non-hermitian
  systems},\ }\href@noop {} {\bibfield  {journal} {\bibinfo  {journal} {arXiv
  preprint arXiv:2401.15000}\ } (\bibinfo {year} {2024})}\BibitemShut {NoStop}%
\bibitem [{\citenamefont {Sirker}\ and\ \citenamefont
  {Kl{\"u}mper}(2005)}]{sirker2005}%
  \BibitemOpen
  \bibfield  {author} {\bibinfo {author} {\bibfnamefont {J.}~\bibnamefont
  {Sirker}}\ and\ \bibinfo {author} {\bibfnamefont {A.}~\bibnamefont
  {Kl{\"u}mper}},\ }\bibfield  {title} {\bibinfo {title} {Real-time dynamics at
  finite temperature by {{DMRG}}: {{A}} path-integral approach},\ }\href
  {https://doi.org/10.1103/PhysRevB.71.241101} {\bibfield  {journal} {\bibinfo
  {journal} {Physical Review B}\ }\textbf {\bibinfo {volume} {71}},\ \bibinfo
  {pages} {241101} (\bibinfo {year} {2005})},\ \Eprint
  {https://arxiv.org/abs/cond-mat/0504091} {arXiv:cond-mat/0504091}
  \BibitemShut {NoStop}%
\bibitem [{\citenamefont {Andraschko}\ and\ \citenamefont
  {Sirker}(2014)}]{andraschko2014}%
  \BibitemOpen
  \bibfield  {author} {\bibinfo {author} {\bibfnamefont {F.}~\bibnamefont
  {Andraschko}}\ and\ \bibinfo {author} {\bibfnamefont {J.}~\bibnamefont
  {Sirker}},\ }\bibfield  {title} {\bibinfo {title} {Dynamical quantum phase
  transitions and the {{Loschmidt}} echo: {{A}} transfer matrix approach},\
  }\href {https://doi.org/10.1103/PhysRevB.89.125120} {\bibfield  {journal}
  {\bibinfo  {journal} {Physical Review B}\ }\textbf {\bibinfo {volume} {89}},\
  \bibinfo {pages} {125120} (\bibinfo {year} {2014})},\ \Eprint
  {https://arxiv.org/abs/1312.4165} {arXiv:1312.4165 [cond-mat]} \BibitemShut
  {NoStop}%
\bibitem [{\citenamefont {Sonner}\ \emph {et~al.}(2021)\citenamefont {Sonner},
  \citenamefont {Lerose},\ and\ \citenamefont
  {Abanin}}]{sonner_influence_2021}%
  \BibitemOpen
  \bibfield  {author} {\bibinfo {author} {\bibfnamefont {M.}~\bibnamefont
  {Sonner}}, \bibinfo {author} {\bibfnamefont {A.}~\bibnamefont {Lerose}},\
  and\ \bibinfo {author} {\bibfnamefont {D.~A.}\ \bibnamefont {Abanin}},\
  }\bibfield  {title} {\bibinfo {title} {Influence functional of many-body
  systems: Temporal entanglement and matrix-product state representation},\
  }\href {https://doi.org/10.1016/j.aop.2021.168677} {\bibfield  {journal}
  {\bibinfo  {journal} {Annals of Physics}\ }\textbf {\bibinfo {volume}
  {435}},\ \bibinfo {pages} {168677} (\bibinfo {year} {2021})}\BibitemShut
  {NoStop}%
\bibitem [{\citenamefont {Lerose}\ \emph
  {et~al.}(2021{\natexlab{b}})\citenamefont {Lerose}, \citenamefont {Sonner},\
  and\ \citenamefont {Abanin}}]{lerose2021}%
  \BibitemOpen
  \bibfield  {author} {\bibinfo {author} {\bibfnamefont {A.}~\bibnamefont
  {Lerose}}, \bibinfo {author} {\bibfnamefont {M.}~\bibnamefont {Sonner}},\
  and\ \bibinfo {author} {\bibfnamefont {D.~A.}\ \bibnamefont {Abanin}},\
  }\bibfield  {title} {\bibinfo {title} {Scaling of temporal entanglement in
  proximity to integrability},\ }\href
  {https://doi.org/10.1103/PhysRevB.104.035137} {\bibfield  {journal} {\bibinfo
   {journal} {Physical Review B}\ }\textbf {\bibinfo {volume} {104}},\ \bibinfo
  {pages} {035137} (\bibinfo {year} {2021}{\natexlab{b}})}\BibitemShut
  {NoStop}%
\bibitem [{\citenamefont {Lerose}\ \emph
  {et~al.}(2021{\natexlab{c}})\citenamefont {Lerose}, \citenamefont {Sonner},\
  and\ \citenamefont {Abanin}}]{lerose2021a}%
  \BibitemOpen
  \bibfield  {author} {\bibinfo {author} {\bibfnamefont {A.}~\bibnamefont
  {Lerose}}, \bibinfo {author} {\bibfnamefont {M.}~\bibnamefont {Sonner}},\
  and\ \bibinfo {author} {\bibfnamefont {D.~A.}\ \bibnamefont {Abanin}},\
  }\bibfield  {title} {\bibinfo {title} {Influence {{Matrix Approach}} to
  {{Many-Body Floquet Dynamics}}},\ }\href
  {https://doi.org/10.1103/PhysRevX.11.021040} {\bibfield  {journal} {\bibinfo
  {journal} {Physical Review X}\ }\textbf {\bibinfo {volume} {11}},\ \bibinfo
  {pages} {021040} (\bibinfo {year} {2021}{\natexlab{c}})}\BibitemShut
  {NoStop}%
\bibitem [{\citenamefont {Foligno}\ \emph {et~al.}(2023)\citenamefont
  {Foligno}, \citenamefont {Zhou},\ and\ \citenamefont
  {Bertini}}]{foligno2023}%
  \BibitemOpen
  \bibfield  {author} {\bibinfo {author} {\bibfnamefont {A.}~\bibnamefont
  {Foligno}}, \bibinfo {author} {\bibfnamefont {T.}~\bibnamefont {Zhou}},\ and\
  \bibinfo {author} {\bibfnamefont {B.}~\bibnamefont {Bertini}},\ }\bibfield
  {title} {\bibinfo {title} {Temporal {{Entanglement}} in {{Chaotic Quantum
  Circuits}}},\ }\href {https://doi.org/10.1103/PhysRevX.13.041008} {\bibfield
  {journal} {\bibinfo  {journal} {Physical Review X}\ }\textbf {\bibinfo
  {volume} {13}},\ \bibinfo {pages} {041008} (\bibinfo {year} {2023})},\
  \Eprint {https://arxiv.org/abs/2302.08502} {arXiv:2302.08502} \BibitemShut
  {NoStop}%
\bibitem [{\citenamefont {Yao}\ and\ \citenamefont {Claeys}(2024)}]{Yao2024}%
  \BibitemOpen
  \bibfield  {author} {\bibinfo {author} {\bibfnamefont {J.}~\bibnamefont
  {Yao}}\ and\ \bibinfo {author} {\bibfnamefont {P.~W.}\ \bibnamefont
  {Claeys}},\ }\href {https://doi.org/10.48550/arXiv.2404.14374} {\bibinfo
  {title} {Temporal {{Entanglement Profiles}} in {{Dual-Unitary Clifford
  Circuits}} with {{Measurements}}}} (\bibinfo {year} {2024}),\ \bibinfo {note}
  {available in: https://arxiv.org/abs/2404.14374},\ \Eprint
  {https://arxiv.org/abs/2404.14374} {arXiv:2404.14374} \BibitemShut {NoStop}%
\bibitem [{\citenamefont {Lu}\ and\ \citenamefont {Grover}(2021)}]{lu2021}%
  \BibitemOpen
  \bibfield  {author} {\bibinfo {author} {\bibfnamefont {T.-C.}\ \bibnamefont
  {Lu}}\ and\ \bibinfo {author} {\bibfnamefont {T.}~\bibnamefont {Grover}},\
  }\bibfield  {title} {\bibinfo {title} {Spacetime duality between localization
  transitions and measurement-induced transitions},\ }\href
  {https://doi.org/10.1103/PRXQuantum.2.040319} {\bibfield  {journal} {\bibinfo
   {journal} {PRX Quantum}\ }\textbf {\bibinfo {volume} {2}},\ \bibinfo {pages}
  {040319} (\bibinfo {year} {2021})}\BibitemShut {NoStop}%
\bibitem [{\citenamefont {Ippoliti}\ \emph {et~al.}(2022)\citenamefont
  {Ippoliti}, \citenamefont {Rakovszky},\ and\ \citenamefont
  {Khemani}}]{ippoliti2022}%
  \BibitemOpen
  \bibfield  {author} {\bibinfo {author} {\bibfnamefont {M.}~\bibnamefont
  {Ippoliti}}, \bibinfo {author} {\bibfnamefont {T.}~\bibnamefont
  {Rakovszky}},\ and\ \bibinfo {author} {\bibfnamefont {V.}~\bibnamefont
  {Khemani}},\ }\href {https://doi.org/10.48550/arXiv.2103.06873} {\bibinfo
  {title} {Fractal, logarithmic and volume-law entangled non-thermal steady
  states via spacetime duality}} (\bibinfo {year} {2022}),\ \Eprint
  {https://arxiv.org/abs/2103.06873} {arXiv:2103.06873} \BibitemShut {NoStop}%
\bibitem [{\citenamefont {Li}\ \emph {et~al.}(2019)\citenamefont {Li},
  \citenamefont {Chen},\ and\ \citenamefont {Fisher}}]{li2019}%
  \BibitemOpen
  \bibfield  {author} {\bibinfo {author} {\bibfnamefont {Y.}~\bibnamefont
  {Li}}, \bibinfo {author} {\bibfnamefont {X.}~\bibnamefont {Chen}},\ and\
  \bibinfo {author} {\bibfnamefont {M.~P.~A.}\ \bibnamefont {Fisher}},\
  }\bibfield  {title} {\bibinfo {title} {Measurement-driven entanglement
  transition in hybrid quantum circuits},\ }\href
  {https://doi.org/10.1103/PhysRevB.100.134306} {\bibfield  {journal} {\bibinfo
   {journal} {Physical Review B}\ }\textbf {\bibinfo {volume} {100}},\ \bibinfo
  {pages} {134306} (\bibinfo {year} {2019})}\BibitemShut {NoStop}%
\bibitem [{\citenamefont {Makri}\ and\ \citenamefont
  {Makarov}(1995)}]{makri1995}%
  \BibitemOpen
  \bibfield  {author} {\bibinfo {author} {\bibfnamefont {N.}~\bibnamefont
  {Makri}}\ and\ \bibinfo {author} {\bibfnamefont {D.~E.}\ \bibnamefont
  {Makarov}},\ }\bibfield  {title} {\bibinfo {title} {Tensor propagator for
  iterative quantum time evolution of reduced density matrices. {{I}}.
  {{Theory}}},\ }\href {https://doi.org/10.1063/1.469508} {\bibfield  {journal}
  {\bibinfo  {journal} {The Journal of Chemical Physics}\ }\textbf {\bibinfo
  {volume} {102}},\ \bibinfo {pages} {4600} (\bibinfo {year}
  {1995})}\BibitemShut {NoStop}%
\bibitem [{\citenamefont {Strathearn}\ \emph {et~al.}(2018)\citenamefont
  {Strathearn}, \citenamefont {Kirton}, \citenamefont {Kilda}, \citenamefont
  {Keeling},\ and\ \citenamefont {Lovett}}]{strathearn2017}%
  \BibitemOpen
  \bibfield  {author} {\bibinfo {author} {\bibfnamefont {A.}~\bibnamefont
  {Strathearn}}, \bibinfo {author} {\bibfnamefont {P.}~\bibnamefont {Kirton}},
  \bibinfo {author} {\bibfnamefont {D.}~\bibnamefont {Kilda}}, \bibinfo
  {author} {\bibfnamefont {J.}~\bibnamefont {Keeling}},\ and\ \bibinfo {author}
  {\bibfnamefont {B.~W.}\ \bibnamefont {Lovett}},\ }\bibfield  {title}
  {\bibinfo {title} {Efficient non-{{Markovian}} quantum dynamics using
  time-evolving matrix product operators},\ }\href
  {https://doi.org/10.1038/s41467-018-05617-3} {\bibfield  {journal} {\bibinfo
  {journal} {Nature Communications}\ }\textbf {\bibinfo {volume} {9}},\
  \bibinfo {pages} {3322} (\bibinfo {year} {2018})}\BibitemShut {NoStop}%
\bibitem [{\citenamefont {Cygorek}\ \emph {et~al.}(2022)\citenamefont
  {Cygorek}, \citenamefont {Cosacchi}, \citenamefont {Vagov}, \citenamefont
  {Axt}, \citenamefont {Lovett}, \citenamefont {Keeling},\ and\ \citenamefont
  {Gauger}}]{cygorek2022}%
  \BibitemOpen
  \bibfield  {author} {\bibinfo {author} {\bibfnamefont {M.}~\bibnamefont
  {Cygorek}}, \bibinfo {author} {\bibfnamefont {M.}~\bibnamefont {Cosacchi}},
  \bibinfo {author} {\bibfnamefont {A.}~\bibnamefont {Vagov}}, \bibinfo
  {author} {\bibfnamefont {V.~M.}\ \bibnamefont {Axt}}, \bibinfo {author}
  {\bibfnamefont {B.~W.}\ \bibnamefont {Lovett}}, \bibinfo {author}
  {\bibfnamefont {J.}~\bibnamefont {Keeling}},\ and\ \bibinfo {author}
  {\bibfnamefont {E.~M.}\ \bibnamefont {Gauger}},\ }\bibfield  {title}
  {\bibinfo {title} {Simulation of open quantum systems by automated
  compression of arbitrary environments},\ }\href
  {https://doi.org/10.1038/s41567-022-01544-9} {\bibfield  {journal} {\bibinfo
  {journal} {Nature Physics}\ }\textbf {\bibinfo {volume} {18}},\ \bibinfo
  {pages} {662} (\bibinfo {year} {2022})}\BibitemShut {NoStop}%
\bibitem [{\citenamefont {Ye}\ and\ \citenamefont {Chan}(2021)}]{ye2021}%
  \BibitemOpen
  \bibfield  {author} {\bibinfo {author} {\bibfnamefont {E.}~\bibnamefont
  {Ye}}\ and\ \bibinfo {author} {\bibfnamefont {G.~K.-L.}\ \bibnamefont
  {Chan}},\ }\bibfield  {title} {\bibinfo {title} {Constructing {{Tensor
  Network Influence Functionals}} for {{General Quantum Dynamics}}},\ }\href
  {https://doi.org/10.1063/5.0047260} {\bibfield  {journal} {\bibinfo
  {journal} {The Journal of Chemical Physics}\ }\textbf {\bibinfo {volume}
  {155}},\ \bibinfo {pages} {044104} (\bibinfo {year} {2021})},\ \Eprint
  {https://arxiv.org/abs/2101.05466} {arXiv:2101.05466 [physics,
  physics:quant-ph]} \BibitemShut {NoStop}%
\bibitem [{\citenamefont {Dowling}\ and\ \citenamefont
  {Modi}(2024)}]{dowling2024a}%
  \BibitemOpen
  \bibfield  {author} {\bibinfo {author} {\bibfnamefont {N.}~\bibnamefont
  {Dowling}}\ and\ \bibinfo {author} {\bibfnamefont {K.}~\bibnamefont {Modi}},\
  }\bibfield  {title} {\bibinfo {title} {Operational {{Metric}} for {{Quantum
  Chaos}} and the {{Corresponding Spatiotemporal-Entanglement Structure}}},\
  }\href {https://doi.org/10.1103/PRXQuantum.5.010314} {\bibfield  {journal}
  {\bibinfo  {journal} {PRX Quantum}\ }\textbf {\bibinfo {volume} {5}},\
  \bibinfo {pages} {010314} (\bibinfo {year} {2024})}\BibitemShut {NoStop}%
\bibitem [{\citenamefont {Park}\ \emph {et~al.}(2024)\citenamefont {Park},
  \citenamefont {Ng}, \citenamefont {Reichman},\ and\ \citenamefont
  {Chan}}]{park2024}%
  \BibitemOpen
  \bibfield  {author} {\bibinfo {author} {\bibfnamefont {G.}~\bibnamefont
  {Park}}, \bibinfo {author} {\bibfnamefont {N.}~\bibnamefont {Ng}}, \bibinfo
  {author} {\bibfnamefont {D.~R.}\ \bibnamefont {Reichman}},\ and\ \bibinfo
  {author} {\bibfnamefont {G.~K.-L.}\ \bibnamefont {Chan}},\ }\bibfield
  {title} {\bibinfo {title} {Tensor network influence functionals in the
  continuous-time limit: Connections to quantum embedding, bath discretization,
  and higher-order time propagation},\ }\href
  {https://doi.org/10.1103/PhysRevB.110.045104} {\bibfield  {journal} {\bibinfo
   {journal} {Phys. Rev. B}\ }\textbf {\bibinfo {volume} {110}},\ \bibinfo
  {pages} {045104} (\bibinfo {year} {2024})}\BibitemShut {NoStop}%
\bibitem [{\citenamefont {Bou-Comas}\ \emph {et~al.}(2024)\citenamefont
  {Bou-Comas}, \citenamefont {Marimón}, \citenamefont {Schneider},
  \citenamefont {Carignano},\ and\ \citenamefont {Tagliacozzo}}]{bou2024}%
  \BibitemOpen
  \bibfield  {author} {\bibinfo {author} {\bibfnamefont {A.}~\bibnamefont
  {Bou-Comas}}, \bibinfo {author} {\bibfnamefont {C.~R.}\ \bibnamefont
  {Marimón}}, \bibinfo {author} {\bibfnamefont {J.~T.}\ \bibnamefont
  {Schneider}}, \bibinfo {author} {\bibfnamefont {S.}~\bibnamefont
  {Carignano}},\ and\ \bibinfo {author} {\bibfnamefont {L.}~\bibnamefont
  {Tagliacozzo}},\ }\href {https://arxiv.org/abs/2409.05517} {\bibinfo {title}
  {Measuring temporal entanglement in experiments as a hallmark for
  integrability}} (\bibinfo {year} {2024}),\ \Eprint
  {https://arxiv.org/abs/2409.05517} {arXiv:2409.05517 [quant-ph]} \BibitemShut
  {NoStop}%
\bibitem [{\citenamefont {Milekhin}\ \emph {et~al.}(2025)\citenamefont
  {Milekhin}, \citenamefont {Adamska},\ and\ \citenamefont
  {Preskill}}]{milekhin2025}%
  \BibitemOpen
  \bibfield  {author} {\bibinfo {author} {\bibfnamefont {A.}~\bibnamefont
  {Milekhin}}, \bibinfo {author} {\bibfnamefont {Z.}~\bibnamefont {Adamska}},\
  and\ \bibinfo {author} {\bibfnamefont {J.}~\bibnamefont {Preskill}},\ }\href
  {https://doi.org/10.48550/arXiv.2502.12240} {\bibinfo {title} {Observable and
  computable entanglement in time}} (\bibinfo {year} {2025}),\ \Eprint
  {https://arxiv.org/abs/2502.12240} {arXiv:2502.12240 [quant-ph]} \BibitemShut
  {NoStop}%
\end{thebibliography}%

\end{document}